\documentclass[preprint1]{aastex701}

\newcommand{\hen}{Hen 2-10}
\usepackage[caption=false]{subfig}
\usepackage{multirow}
\usepackage{booktabs}
\usepackage[normalem]{ulem}
\usepackage{xcolor}

\shortauthors{Dalsin et al.}

\begin{document}
%
%
%
%
%
%

\newcommand{\PI}{Principal Investigator (PI)}
\newcommand{\PFT}{Proposal finder tool (PFT)}
\newcommand{\SRT}{Science Review Panel (SRP)}
\newcommand{\TAC}{Telescope Allocation Committee (TAC)}
\newcommand{\TTA}{Telescope Time Allocation (TTA)}

\newcommand{\vlatext}{The Karl G. Jansky Very large Array is an instrument of the National Radio Astronomy Observatory (NRAO). The NRAO is a facility of the National Science Foundation, operated under cooperative agreement with Associated Universities, Inc.}

\newcommand{\hco}{HCO$^+$}
\newcommand{\otz}{(1-0)}
\newcommand{\tto}{(2-1)}
\newcommand{\ttt}{(3-2)}
\newcommand{\twco}{$^{12}$CO}
\newcommand{\thco}{$^{13}$CO}
\newcommand{\palpha}{Pa$\alpha$}
\newcommand{\ohm}{12 + log(O/H) = }
\newcommand{\ohz}{Z$_\odot$}
\newcommand{\xco}[1]{#1 $\times$ 10$^{20}$ cm$^{-2}$ (K km s$^{-1}$)$^{-1}$}


\newcommand{\PDR}{Photodissociation Region}
\newcommand{\RomanNumeralCaps}[1]{\MakeUppercase{\romannumeral #1}}
\newcommand{\bext}{\textbf{B}$_{\textrm{ext}}$}
\newcommand{\mhii}{\textrm{H}\,\textsc{ii}}
\newcommand{\mPhii}{P^{\textrm{H}\,\textsc{ii}}_{\textrm{th}}}
\newcommand{\Phii}{$P^{\textrm{H}\,\textsc{ii}}_{\textrm{th}}$}
\newcommand{\mPpdr}{P^{\textrm{PDR}}_{\textrm{th}}}
\newcommand{\Ppdr}{$P^{\textrm{PDR}}_{\textrm{th}}$}
\newcommand{\PAH}{Polycyclic aromatic hydrocarbons (PAH)}

\newcommand{\VLASS}{VLA Sky Survey (VLASS)}
\newcommand{\VLITE}{VLA Low-band Ionosphere and Transient Experiment (VLITE)}
\newcommand{\CGPS}{Canadian Galactic Plane Survey (CGPS)}
\newcommand{\NRAO}{National Radio Astronomy Observatory}
\newcommand{\CASAfootnote}{\footnotesize{See the NRAO Jansky VLA tutorial ``EVLA Continuum Tutorial 3C391'' (http$://$casaguides.nrao.edu$/$index.php$?$title$=$EVLA$_{-}$Continuum$_{-}$Tutorial$_{-}$3C391)}}
\newcommand{\WISE}{Wide-field Infrared Survey Explorer (WISE) from the IPAC All-Sky Data Release}
\newcommand{\CASA}{Common Astronomy Software Applications (CASA)}
\newcommand{\ALMA}{Atacama Large Millimeter/submillimeter Array (ALMA)}
\newcommand{\ASKAP}{Australian Square Kilometre Array Pathfinder (ASKAP)}
\newcommand{\POSSUM}{Polarisation Sky Survey of the Universe's Magnetism (POSSUM)}
\newcommand{\EMU}{Evolutionary Map of the Universe (EMU)}
\newcommand{\AIPS}{Astronomical Image Processing System (AIPS)}
\newcommand{\HST}{Hubble Space Telescope(HST)}
\newcommand{\STScI}{Space Telescope Science Institute (STScI)}
\newcommand{\JWST}{James Webb Space Telescope (JWST)}

\newcommand{\pdr}{photodissociation region}
\newcommand{\ghz}{$\;$GHz}
\newcommand{\pk}{$P/k_b$}


\newcommand{\fdf}{\textit{F}$(\phi)$}
\newcommand{\I}{\textit{I}}
\newcommand{\Q}{\textit{Q}}
\newcommand{\U}{\textit{U}}
\newcommand{\Pol}{\textit{P}}
\newcommand{\QU}{\textit{QU}}
\newcommand{\thetaP}{$\theta_{\textrm{{\scriptsize peak}}}$}
\newcommand{\maxL}{ $\mathscr{L}_{\textrm{{\scriptsize max}}}$}
\newcommand{\chilam}{$\chi(\lambda^{2})$}


\newcommand{\Alfven}{Alfv\'{e}n }
\newcommand{\Alfvenic}{Alfv\'{e}nic }
%
%
\newcommand{\strom}{Str\"{o}mgren }
\newcommand{\hi}{H\,{\sc i}} 
\newcommand{\HII}{H\,{\sc ii} } 
\newcommand{\ddeg}{\ensuremath{^{\circ}}} 
\newcommand{\arcminn}{\ensuremath{^{\prime}}}
\newcommand{\arcsecc}{\ensuremath{^{\prime\prime}}}
\newcommand{\lsi}{LSI +61\ddeg303}
\newcommand{\sunn}{\ensuremath{\odot}}
\newcommand{\Msun}{\ensuremath{\textrm{M}_{\odot} }} 
\newcommand{\MLu}{\textrm{M}$_{\odot}$ yr$^{-1}$}
\newcommand{\kms}{km s$^{-1}$}
\newcommand{\infinity}{\ensuremath{\infty}}
\newcommand{\ML}{\ensuremath{\dot{M}}} 
\newcommand{\Lsunn}{\ensuremath{L_{\odot}}}  
\newcommand{\radm}{rad m$^{-2}$}
\newcommand{\MLM}{\dot{M}}
\newcommand{\blos}{B$_{los}$}
\newcommand{\cm}{cm$^{-3}$}
\newcommand{\plam}{\textit{P($\lambda^2$)}}
\newcommand{\fsimp}{\textit{Faraday simple}}
\newcommand{\fcomp}{\textit{Faraday complex}}
\newcommand{\fphi}{\textit{F($\phi$)}}
\newcommand{\lamsq}{$\lambda^2$}
\newcommand{\rmclean}{\textsc{rmclean}}
\newcommand{\clean}{\textsc{clean}}
\newcommand{\immath}{\textsc{immath}}
\newcommand{\polcal}{\textsc{polcal}}
\newcommand{\equa}[1]{Equation (\ref{eq:#1})}
\newcommand{\equas}[2]{Equations (\ref{eq:#1}) and (\ref{eq:#2})}
\newcommand{\about}{$\sim$}
\newcommand{\mircon}{$\mu$m}

\defcitealias{Kobulnicky:1995}{K95}

\title{Revisiting the Gas Dynamics of Henize 2-10: Possible Drivers of the Starburst}

\author[0009-0003-3265-5589]{Josephine M. Dalsin}
\affiliation{National Radio Astronomy Observatory, Charlottesville, Virginia}
\email{jdalsin@nrao.edu}

\author[0000-0002-7408-7589]{Allison H. Costa}
\affiliation{National Radio Astronomy Observatory, Charlottesville, Virginia}
\email{acosta@nrao.edu}

\author[0000-0002-4663-6827]{Remy Indebetouw}
\affiliation{Department of Astronomy, University of Virginia, Charlottesville, Virginia}
\affiliation{National Radio Astronomy Observatory,
Charlottesville, Virginia}
\email{rindebet@nrao.edu}

\author[0000-0001-8348-2671]{Kelsey E. Johnson}
\affiliation{Department of Astronomy, University of Virginia, Charlottesville, Virginia}
\email{kej7a@virginia.edu}

\author[0000-0002-4013-6469]{Natalie O. Butterfield}
\affiliation{National Radio Astronomy Observatory,
Charlottesville, Virginia}
\email{nbutterf@nrao.edu}

\author[0000-0002-7532-3328]{Sabrina Stierwalt}
\affiliation{Physics Department, 1600 Campus Road, Occidental College, Los Angeles, California}
\email{sabrina@oxy.edu}

\begin{abstract}
The triggers of starburst episodes are a key component to our understanding of the baryon cycle in galaxies. Galaxy mergers are a commonly suggested catalyst for starbursts, but once the galaxies coalesce into a single kinematically disturbed system, their merger history can be difficult to assess. This is particularly true for dwarf galaxies, which are expected to dominate the merger rate at all redshifts due to their large numbers. One such dwarf galaxy undergoing an enigmatic starburst episode is Henize 2-10, which appears to be isolated. Possible scenarios that might have caused the starburst episode include a previous merger or stochastic processes within the galaxy itself, such as self-regulation via feedback processes.  We present new VLA 21-cm observations and unpublished archival CARMA CO data to investigate the dynamical state and star formation activity in the galaxy. We do not detect an \hi{} tail consistent with the structure reported by \citet{Kobulnicky:1995}, which was suggested as evidence for a merger or interaction, but rather these new observations indicate an extended \hi{} distribution. We also find that the \hi{} appears dynamically decoupled from an extended CO feature (inferred to be a tidal tail in previous work), suggesting large-scale dynamical processes of some type are affecting the gas in this system.  We provide a meta-analysis of available results to enhance our understanding of what might be triggering the starburst episode in Henize 2-10, and speculate that the large CO feature could be falling into the galaxy and potentially trigger starburst activity.

\end{abstract}


\section{Introduction\label{sec:intro}}

Starbursts represent an intense yet unsustainable phase of star formation in galaxies, with high star formation rates (SFRs) and star formation efficiencies \citep[e.g.,][]{Rieke:1980,Tenorio-Tagle:2003,Jaskot:2015}. Starbursting galaxies can also host extreme modes of star formation \citep[e.g.,][]{deGrijs:2001,OConnell:1995,Watson:1996,Goddard:2010,Kruijssen:2012,Pfeffer:2019,Lahen:2020}. One such mode is the formation of super star clusters (SSCs), which are remarkable sites of massive star formation. These clusters have been actively studied since the time of early HST observations, and exhibit stellar densities exceeding 10$^4$ stars per pc$^{3}$ in their cores and masses often surpassing 10$^{4.5}$ \Msun{} \citep{Whitmore:1995, Johnson99, Kobulnicky:1999, Johnson:2000,Portegies:2010}. 
Identifying the driver of the starburst activity can provide valuable insight into the environments that foster such extreme stellar products.

External mechanisms, such as mergers or interactions, have been both observed and simulated as causes of starburst activity \citep[e.g.,][]{Barnes:1991,Jog:1992,Mihos:1996,Noeske2001, Ostlin2001, Pustilnik2001, Bournaud:2008, Kim:2009, Moreno:2015, Bradford:2015,Stierwalt2015,Pearson:2016,Hopkins:2018}. However, observational evidence for mergers, particularly in advanced stages, can be difficult to identify, as the signatures of the interaction become less clear, leaving behind what appears to be a single disturbed system \citep{Toomre1972, Bekki2008,Pearson:2018}. The dynamical complexity of mergers can make them challenging to diagnose. On one hand, gas can also remain extended for longer periods \citep[][]{ThuanMartin1981,Hibbard2000,Hibbard2001b,Thuan2004,Ashley2013, Ashley2017}, retaining important clues about the system's dynamical history in the form of tidal features such as bridges, shells, loops, plumes, and tails \citep{Arp:1966,Schombert:1990,Kim:2009,Duc:2013,Privon:2013}. On the other hand, gas can also flow back into the merging system to fuel the starburst, rapidly erasing many of the original merger imprints \citep[e.g.,][]{Hopkins:2008d,Hopkins:2009,Sacchi:2024}. The extent to which tell-tale signs of mergers and interactions seen in massive galaxies can be expected for dwarf galaxies is also unclear \citep[e.g.][]{Stierwalt2015}; their lower masses, weaker tides, and lack of stabilizing structures such as spiral arms result in fundamentally different dynamical signatures. For example, late stage dwarf-dwarf mergers may be associated with tidal debris that are stubbier and less tail-like \citep{KadoFong:2020}. 

Internal mechanisms, such as radiation, supernovae, or stellar winds, are also thought to trigger bursts of star formation in dwarf galaxies. \citet{Stinson:2007} simulate feedback in dwarf galaxies and find that stellar winds and supernovae quench star formation processes by disrupting the gas, then when star formation ceases, the gas cools and accretes, initiating a new episode of star formation. This gas recycling process is often referred to as a galactic fountain \citep[][]{Shapiro:1976,Chisholm:2016}. Even in isolated dwarf galaxies, the simulations of \citet{Stinson:2007} indicate it is possible to produce  episodic star formation via this process. In dwarf starbursts, \citet{McQuinn:2019} and \citet{Marasco:2023} find that the majority of material expelled by stellar feedback remains bound to the galaxy, which would allow it to re-accrete. Internal feedback can rapidly disrupt the \hi{} gas of a galaxy, which can manifest as a patchy \hi{} distribution and disorganized kinematics. In dwarf galaxies though, the short dynamical timescales allow the \hi{} reservoirs to return a regular, organized distribution after several tens of Myr \citep{Stinson:2007,Bernstein-Cooper:2014,Rey:2020,Rey:2022,Rey:2024}.

Extended \hi{} envelopes are characteristic of dwarf galaxies \citep[][]{Swaters:2002,Stierwalt2015,Pearson:2016}, and because the host's potential well is shallow, these envelopes are readily disturbed. During interactions or mergers, the resulting tidal debris are typically \hi{}-rich \citep[e.g.,][]{Yun:1994,Hibbard2001a,Martinsson:2013,Privon:2013,Eibensteiner:2023,Eibensteiner:2024}. Conversely, galactic winds can form \hi{} gas into extended structures that might be mistaken for tidal features \citep{Meurer:1998}. Examining 18 star-bursting dwarfs, \citet{Lelli:2014b,Lelli:2014} showed that systems whose outer-disk orbital time is comparable to the age of the most recent star-formation episode display relatively symmetric \hi{} morphologies, whereas strongly asymmetric \hi{} distributions are linked to more recent bursts. Simulations indicate that the \hi{} distribution tracks star-formation history \citep{Rey:2020,Rey:2022,Rey:2024}, but distinguishing between internally and externally driven mechanisms remains difficult, particularly in the absence of a clear perturber. %

\subsection{Case Study: Henize 2-10\label{sec:hen}}

Henize 2-10 (Hen 2-10) is a nearby (9 Mpc; \citealt{Vacca:1992}) starbursting blue compact dwarf (BCD) galaxy that hosts several hundred Wolf-Rayet stars as well as many SSCs \citep{Conti1991,Vacca:1992,Conti:1994, Allen:1998,Kobulnicky:1999,Johnson:2000}. Estimates of the SFR vary between 0.76 -- 1.9 \MLu{} \citep[][]{Madden2013,Reines:2011,Cresci:2017,Marasco:2023}. \citet{Reines:2011} estimate the stellar mass (M$_{\ast}$) as $3.7 \times 10^9$ \Msun; \citet{Nguyen2014} find a higher mass of (9.8 $\pm$ 3.0) $\times 10^9$ \Msun{} within 100\arcsec{}. The specific star formation rate (sSFR = SFR / M$_{\ast}$; \citealt{Leroy:2019}) is $0.19-0.51 \text{ Gyr}^{-1}$ for these ranges of SFR and M$_{\ast}$; for comparison, the Milky Way's sSFR is $\sim$0.02 Gyr$^{-1}$ \citep{Chomiuk:2011,Kennicutt:2012}.

\begin{figure}[htb!]
    \centering
    \subfloat{\includegraphics[width=0.7\linewidth]{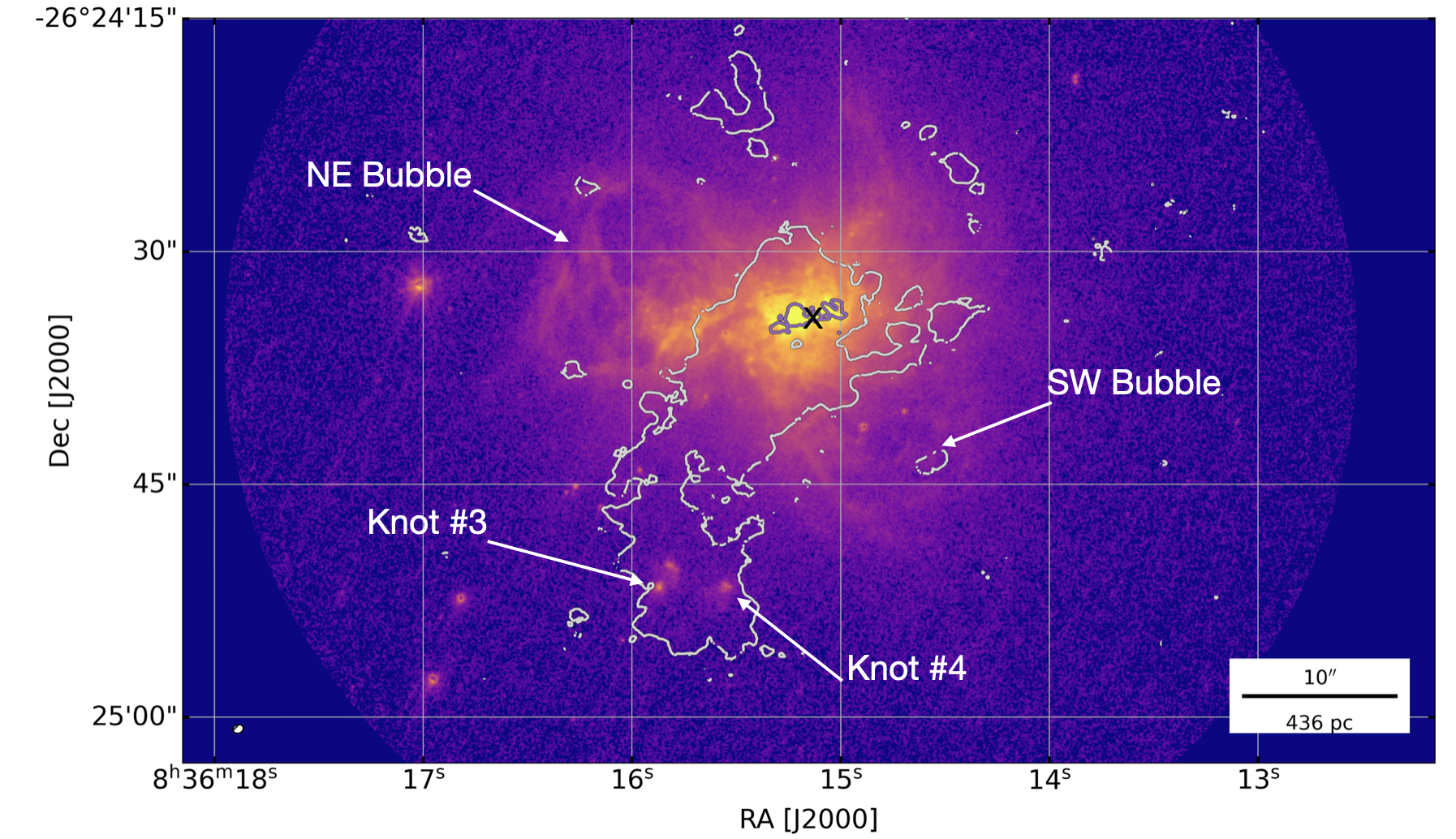}\label{fig:f555FinderA}}
    \quad
\subfloat{\includegraphics[width=0.7\linewidth]{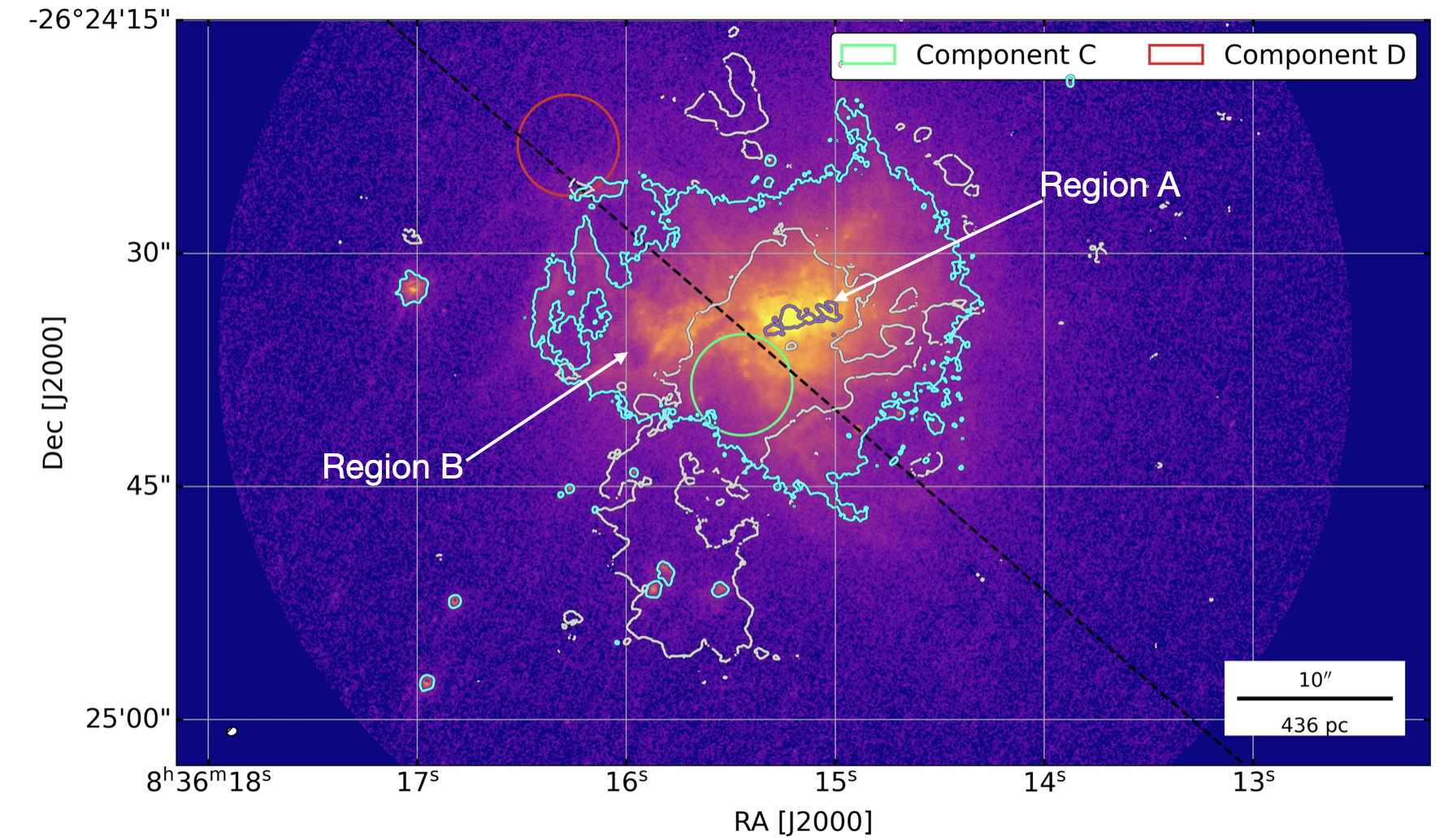}\label{fig:f555FinderB}}
    \caption{Finder charts for Hen 2-10. The raster is a Hubble Legacy Archive HST/WFPC2 H$\alpha$ (F656N) map originally published in \citet{Johnson:2000}. In both panels, the purple contour shows the VLA 33 GHz continuum, which traces free-free emission from the embedded SSCs, from \citet{Costa:2021} at 5$\times$ their rms of 3.12 $\mu$Jy beam$^{-1}$. The gray contour shows this work's \twco{(1-0)} moment 0 map at 4$\sigma_{CO}$; the CO beam is shown with the white ellipse in the lower left. \textit{Top.} The NE and SW outflows and the locations of the two star-forming knots in the SE are annotated \citep{Mendez:1999}. The black `x' on centered in the 33 GHz contours represents the position of the central AGN.
    \textit{Bottom.} The nominal positions of Regions A and B are annotated, and components C and D are shown with circles the size of the \citealt{Kobulnicky:1995} OVRO synthesized beam in green and red, respectively.  The cyan contour shows a representative ``footprint'' of the continuum subtracted F656N map. This contour was chosen to highlight NE and SW bubbles and to distinguish, for illustrative purposes, the CO ``tail'' from the CO main body, the latter of which is enclosed by the cyan contour. The black dashed line is drawn to guide the eye; it passes through the centers of the NE and SW bubbles at a PA = 49\ddeg{} from the +y axis.
    }
    \label{fig:f555Finder}
\end{figure}

The starburst activity is primarily in two regions, noted in Figure \ref{fig:f555Finder} as Region A and Region B, following the nomenclature of \citet{Vacca:1992}. Region A, nominally centered at (RA, Dec.)\footnote{All coordinates are given in the J2000 epoch.} = (08:36:15.2, -26:24:34.3), corresponds to the optical center of the galaxy, where a low-luminosity black hole is believed to reside \citep{Reines:2011,Reines:2016,Riffel:2020}. This region contains both adolescent, optically exposed SSCs and deeply embedded (A$_v$~$\sim$~2–16) natal SSCs that have not yet significantly disrupted their birth environments \citep{Kobulnicky:1999,Johnson:2000,Johnson:2003,Cabanac:2005,Costa:2021}. Region B, at (RA, Dec.) = (08:36:15.7, -26:24:35.0), lies approximately 8\arcsec{} (305 pc at a distance of 9 Mpc) east of Region A and hosts massive stars consistent with B-type stellar populations \citep{Johansson:1987,Corbin:1993,Johnson:2000,Nguyen2014}. Using H$\alpha$ and X-ray observations, \citet{Mendez:1999} reported kiloparsec-scale bipolar outflows extending to the NE and SW of the starburst regions, which they suggested may be driven by supernovae and stellar winds \citep[see also][]{Johnson:2000,Cresci:2017}. In their HST H$\alpha$ data, \citet{Johnson:2000} estimated that the velocity of the outflow is sufficient to escape the gravitational potential well of Hen~2-10.

The intense and concentrated star formation in Hen 2-10 has motivated a variety of investigations to understand the environments that promote SSC formation \citep[e.g.,][]{Johnson:2003,Cabanac:2005,Johnson:2018,Gim:2024a,Gim:2024b}, as SSCs are not ubiquitous objects in the local universe. It is thought that extreme pressures \(P/k_B > 10^7 \textrm{ K cm$^{-3}$} \) across cubic parsecs in the interstellar medium are needed to form SSCs, and merging systems are one such environment that can induce the necessary pressures \citep{Jog:1992,Elmegreen:2002,Johnson:2015,Finn:2019,Keller:2020,Horta:2021}. Whether or not Hen 2-10 is a merger has been a topic of ongoing speculation in the literature for many decades. Early optical and single-dish 21-cm observations by \citet{Allen1976} revealed an extended \hi{} envelope surrounding two central optical peaks. \citet{Johansson:1987} interpreted the two peaks as being two dwarf galaxies either merging or undergoing a close encounter. \citet{Corbin:1993}, on the other hand, proposed that Hen 2-10 is not a merging system but an isolated system undergoing stochastic star formation.

From 21-cm and CO observations, \citet[][hereafter K95]{Kobulnicky:1995} proposed two possible scenarios for Hen 2-10’s starburst activity: a moderately advanced merger between dwarf galaxies or sustained star formation fueled by accreting gas clouds. Using both \hi{} and CO observations, \citetalias{Kobulnicky:1995} identified a prominent molecular extension in their Owens Valley Radio Observatory (OVRO) \twco{(1-0)} data, which extends approximately 15\arcsec{} southeast of Region A and was interpreted as a tidal tail (see Figure \ref{fig:f555Finder}). We refer to this structure as the CO ``tail'' hereafter as the true nature of the molecular material is not known. The CO ``tail'' had previously been detected by \citet{Baas:1994} in the \twco{(3-2)}, \twco{(2-1)}, and \twco{(1-0)} transitions, although it was only partially resolved in their single-dish observations. From interferometric 21-cm observations, \citetalias{Kobulnicky:1995} also identified an \hi{} counterpart to the CO ``tail'', along with fainter \hi{} emission extending out to $\sim$ (RA, Dec.) = (08:36:18.0, $-$26:25:35.7), southeast of the starburst regions (see their Figure 12). They similarly interpreted this extended \hi{} emission as a tidal feature. \citetalias{Kobulnicky:1995} then suggest that if those gas features are tidal, that their existence supports the late interaction scenario. 

In the second scenario, \citetalias{Kobulnicky:1995} hypothesized that starburst activity could be sustained by the accretion of small gas clouds, with masses on the order of 10$^5$--10$^6$ \Msun{}. This second hypothesis was motivated by their \hi{} mass estimate, which was only 63\% of the value reported in earlier single-dish observations by \citet{Allen1976}. Later single dish \hi{} studies by \citet{Sauvage1997} and \citet{Meyer:2004} are consistent with the measurement of \citet{Allen1976}, further suggesting that a significant amount of \hi{} gas could be dispersed in more diffuse structures not detected in the \citetalias{Kobulnicky:1995} data. \citetalias{Kobulnicky:1995} proposed that a kinematically distinct CO component, located between Region A and the CO ``tail'', may be an example of a small cloud fueling the starburst.  It is redshifted by +30~\kms{} relative to the systemic velocity\footnote{All velocities are reported in the LSRK frame unless explicitly noted otherwise.} (v$_{sys,Helio}$) of (873 $\pm$ 10)~\kms{} and has a molecular mass of approximately 4 $\times$ 10$^6$ \Msun{} \citepalias{Kobulnicky:1995}. The position of the cloud is labeled as ``component C'' in Figure \ref{fig:f555Finder} near (RA, Dec.) = (08:36:15.5, -26:24:38.5). \citetalias{Kobulnicky:1995} also found a second kinematically distinct CO feature northeast of Region B that is redshifted by +60 \kms{} relative to v$_{sys,Helio}$. We refer to this feature as ``component D'', it is shown in the upper left of Figure \ref{fig:f555Finder} at (RA, Dec.) = (08:36:16.5, -26:24:23.1). While initially attributed to noise, \citet{Vanzi:2009} detected emission at this location in their \twco{(3-2)} Atacama Pathfinder EXperiment (APEX) observations. \citet{Vanzi:2009} estimated the molecular mass of this feature to be M$_{H2}$ = 1.8 $\pm$ 0.5 $\times$ 10$^7$ \Msun{} (scaled to a distance of 9 Mpc and an CO-to-$H_2$ conversion factor 3 $\times$ 10$^{20}$ cm$^{-2}$ (K \kms)$^{-1}$) and suggested that it could be a detached cloud that is starting to form stars.

While earlier studies explored the large scale CO and \hi{} features, more recent investigations have concentrated on the molecular content near Regions A and B. \citet{Johnson:2018} obtained observations with Atacama Large Millimeter/submillimeter Array (ALMA) to map HCN(1-0), HNC(1-0), HCO$^+$(1-0) and CCH(1-0) with arcsecond resolution, finding that dense compact clouds are associated with the natal SSCs. They also identified pre-cursor clouds southeast of Region A and suggested that the molecular material has the potential to form SSCs similar to the current population of Region A. \citet{Johnson:2018} identified a bright CCH(1-0) feature that is partly coincident with component C and noted that if component C is an infalling gas cloud, then the CCH feature highlights the intersection of the cloud and the stellar radiation field from the center regions, which would provide circumstantial evidence for an infalling molecular cloud. 

With sub-arcsecond resolution, \citet{Beck:2018} mapped the center of Hen 2-10 in \twco{(3-2)} with ALMA, and their maps resolved the molecular emission seen by \citetalias{Kobulnicky:1995} into filamentary structures, which they argued are potentially supplying gas to the starburst regions. They proposed that a filament north of Region B could represent a remnant of a larger system that has since dispersed or been depleted after fueling the starburst activity. Their FOV does not extend far enough to include component D, however, and therefore that work is unable to investigate if a connection exists. They also argued a redshifted filament west of Region A (their ``western filament'') is supplying molecular material to the starburst region. 
 
In the \citet{Santangelo_Testi_Gregorini_Leurini_Vanzi_Walmsley_Wilner_2009} \twco{(2-1)} data from the Submillimeter Array (SMA), they identified compact CO clumps associated with young SSCs in the center region; however, the CO ``tail'' is $<$ 2$\sigma$ in their data. \citet{Imara:2019} mapped the CO ``tail'' and main body with the \twco{(1-0)} transition at sub-arcsecond resolution with ALMA. In the center of the galaxy, the \twco{(1-0)} gas shows similar velocity structures and morphologies to the \twco{(3-2)} gas of \citet{Beck:2018} (see Figure 4 of \citealt{Imara:2019} and Figure 6 of \citealt{Beck:2018}). \citet{Imara:2019} found compact clumps in the CO ``tail'' and suggested that the clumps may potentially form SSCs but have not yet done so. However in their Nordic Optical Telescope H$\alpha$ observations, \citet{Beck:1999} and \citet{Mendez:1999} found two potential star forming clumps, referred to as Knots\footnote{We use ``Knots'' here for consistency with \citet{Mendez:1999}, but we note that this term is also used in other studies \citep[e.g.,][]{Kobulnicky:1999,Johnson:2003,Costa:2021} of Hen 2-10 to discuss radio clumps at the center of the galaxy that are unassociated with these H$\alpha$ clumps.} \#3 and \#4, at the southern end of the CO ``tail''. 

In this work, we present new interferometric 21-cm observations of Hen 2-10 and previously unpublished Combined Array for Research in Millimeter-wave Astronomy (CARMA) \twco{(1-0)} and \twco{(2-1)} observations to characterize connections between the \hi{} and CO gas. Sections \ref{sec:HIobs} and \ref{sec:COobs} detail the observational and archival data utilized for this study, which includes combining the CARMA data with other archival data observed with the SMA and ALMA. Section \ref{sec:analysis} describes the data reduction process, data products, and analysis, which include moment maps, line profiles, and position velocities slices. In Section \ref{sec:results}, we present the results, and in Section \ref{sec:discuss}, we discuss these results in a multi-wavelength context to examine the potential external and internal drivers of the starbursting regions in Hen 2-10.

\section{Observations} \label{sec:obs}

\subsection{New VLA HI Observations} \label{sec:HIobs}

\begin{deluxetable}{lcc}
\tablenum{1}
\tablecaption{New VLA Observations\label{tab:VLAobs}}
\tablewidth{0pt}
\tablehead{\colhead{} & \colhead{Continuum} & \colhead{Spectral Line}}
\startdata
VLA Project Code & \multicolumn{2}{c}{21A-198}\\
Number of Schedule blocks & \multicolumn{2}{c}{2} \\
Date of Observations & \multicolumn{2}{c}{2021-07-11 $\&$ 2021-07-20}\\
Total integrated time on source (hr) & \multicolumn{2}{c}{3.2}\\
VLA Configuration & \multicolumn{2}{c}{C-array} \\
Center Frequency of IFs\tablenotemark{a} (GHz) & 1.25, 1.75 & 1.415\\
Number of Frequency Channels & 1920 & 1024\\
Channel width (kHz) & 500 & 7.8\\
Bandwidth (MHz) & 960 & 8 \\
Channel width (\kms) & \nodata & 1.65 \\
Bandwidth (\kms) & \nodata & 1690\\
\enddata
\tablenotetext{a}{Two intermediate frequency (IF) bands were placed at the frequencies listed in the second column for the continuum observation. Within the 1.25 GHz IF, a single narrow subband was centered at frequency in the third column for the spectral line observation. The spectral line and continuum data acquisition were simultaneous.}
\end{deluxetable}

We observed Hen 2-10 with the Karl G. Jansky Very Large Array\footnote{\vlatext} (VLA) in C-array at L band (1--2 GHz).  The observation configuration consisted of one subband, centered at the \hi{} line with a spectral resolution of 7.8 kHz (1.65 \kms) and a total bandwidth of 8 MHz (1698 \kms). We obtain simultaneous continuum observations between 1238 -- 2262 MHz and 738 MHz -- 1762 MHz, divided between 14 subbands of 64 MHz each, for a total bandwidth of 950 MHz. Table \ref{tab:VLAobs} shows a summary of the observational parameters. We used 3C147 as the bandpass and the flux calibrator. We also observed 3C286 as an additional check on the flux density calibration. Observations of the complex gain calibrator, J0837-1951, were interleaved every 15 minutes with the target.

\subsection{CARMA CO Data} \label{sec:COobs}

We present previously unpublished Combined Array for Research in Millimeter-wave Astronomy (CARMA) data from project c1274 (PI: Stierwalt, S). The observations of Hen 2-10 were performed in 2 configurations.
Hen 2-10 was observed with CARMA's C-array (8.2-97.6 k$\lambda$) between 9 and 14 April 2014 for a total of 488 minutes on source.  Spectral windows, with a total bandwidth of 124.22 MHz and a channel width of 781.25 kHz, were placed on $^{12}$CO(1-0), $^{12}$CO(2-1), $^{13}$CO(1-0), and C$^{18}$O(1-0). Equivalently, the total velocity span per 124.22 MHz window is $\sim$161~\kms{} for the 1$\rightarrow$0 transitions and 323 \kms{} for the \twco{(2-1)} transition; the channel widths are $\sim$1~\kms{} for the 1$\rightarrow$0 transitions and $\sim$2~\kms{} for the \twco{(2-1)} transition. Between 14 and 18 May, Hen 2-10 was observed in D-array (5-80.2 k$\lambda$) for a total of 336 minutes on source with the same spectral configuration. For both datasets, the bandpass response was calibrated using J0927+390, the amplitude using 3C84, and time-varying gain using J0730-116.


\begin{deluxetable*}{ccccccccc}
\tablecaption{Summary of Data and Final Data Products\label{tab:obs}}
\tabletypesize{\small}
\tablehead{
\colhead{Project} & \colhead{line} &\colhead{$\nu$} & \colhead{\texttt{clean}} & \colhead{$\delta$v} & \colhead{$\theta_{res}$} & \colhead{$\theta_{LAS}$} & \colhead{$\sigma$}  & \colhead{$\sigma_{N(HI)}$}\\
\colhead{} & & \colhead{(GHz)} & \colhead{\textit{robust}} & \colhead{(\kms)} &\colhead{(\arcsecc)} & \colhead{(\arcsecc)}  & \colhead{(mJy beam$^{-1}$)} & \colhead{(cm$^{-2}$)}
}
\startdata
VLA 21A-198 & HI & 1.42 & 0.5 & 10 &30 $\times$ 13   & 970  & 0.5 & 2.5 $\times$ 10$^{19}$\\
\\
CARMA c1274 + \\ALMA 2016.1.0027.S & $^{12}$CO(1-0) & 115  & 1.5  & 3 & 0.69 $\times$ 0.55   & 112  & 1.4\tablenotemark{a} & \nodata\\
\\
CARMA c1274 + \\SMA 2008B-S043& \twco{(2-1)} & 230  & 2.0 & 10 & 4.4 $\times$ 2.6 & 56 & 42.4 & \nodata\\
\enddata
\tablecomments{Columns 1 -- 3 give the Project Code, the transition, and nominal rest frequency, respectively. Column 4 gives the weighting parameter used in imaging in CASA, and column 5 lists the velocity resolution per bin in the final data products. Columns 6 -- 8 give the synthesized beam, largest angular scale, and the MAD STD statistic per velocity bin, respectively, in the final maps. Column 9 gives the \hi{} column density rms, calculated with Equation \ref{eq:column} over a 10 \kms{} channel. In row 2, the final data product includes CARMA and ALMA observations of the \twco{(1-0)} transition. In row 3, the final data product includes CARMA and SMA observation of the \twco{(2-1)} transition.}
\tablenotetext{a}{This is a lower uncertainty; see Section \ref{sec:codatared}.}
\end{deluxetable*}

\section{Data Reduction, Data Products, and Analysis} \label{sec:analysis}
\subsection{HI Data Reduction\label{sec:vladataredu}}
The VLA project 21A-198 \hi{} data reduction and imaging were done with the NRAO's Common Astronomy Software Applications (CASA; \citealt{CASA}) version 6.4. Each observing epoch was reduced independently. We employed systematic flagging  procedures (e.g., ``Quack'') before visually inspecting the data to manually remove data corrupted by radio frequency interference (RFI).  In the narrow subband centered on the redshifted 21-cm line, there was persistent RFI at 1414.0 -- 1414.2 MHz across all scans and  intermittent interference at 1414.0 -- 1414.2 MHz and 1418 -- 1420.0 MHz on 2 of the 11 scans of the complex gain calibrator. The effected channels were excised prior to calibration. We used standard calibration practices to reduce the data, including determining complex gain solutions, bandpass solutions, and the flux density scale using 3C147. 

The calibrated target data were split into a new measurement set for each observing epoch and then were individually processed with the CASA task \texttt{statwt}, which computes the variance, weighted by exposure time, for real and imaginary parts of the visibilities. These weights were then applied to the visibilities. The two weighted measurements sets were then imaged together using the CASA \texttt{tclean} task with a \textit{specmode} parameter set to `cube' to produce a spectral cube. In this mode, we set the width to be 10~\kms{} and the rest frequency to the \hi{} transition (1.4204058 GHz). The \textit{gridder} was set to \textit{wproject}\footnote{See \url{https://casadocs.readthedocs.io/en/stable/notebooks/synthesis_imaging.html} or Lecture 19, Section 2 in \citet{nraowhitebook}.}, with nterms equal to 26 to account for non-coplanar baselines as is appropriate for the large field of view at 1.42 GHz and the extended nature of Hen 2-10. We set the \textit{pblimit} to a small, negative value to ensure the full extent of the primary beam was included in the imaging field of view and set the \textit{deconvolver} to `hogbom'. The restoring beam was set to `common', to ensure a consistent synthesized beam across all channels. \textit{Briggs} weighting with a \textit{robust} of 0.5 was used to improve the spatial resolution of the data at a modest cost in sensitivity. The number of iterations (niter) was set to 0 and the interactive mode was disabled. This produces a ``dirty'' image, which is sufficient for an initial inspection to identify line-free channels for continuum subtraction. 

We inspected the data using the Cube Analysis and Rendering Tool for Astronomy (CARTA; \citealt{CARTA}) to identify line-free channels for continuum subtraction. Conservatively, we selected channels at velocities less than 700~\kms{} or greater than 990~\kms{} as line free channels and used the CASA task \texttt{uvcontsub} to subtract the continuum from each measurement set in the visibility domain. This task creates a new measurement set containing the continuum-subtracted visibilities. These two new measurement sets were then imaged together using the inputs described above, this time with interactive mode enabled and the iteration count set to 10000.

Once the \texttt{clean}ing process initialized, each channel in the spectral cube was visually inspected and conservative masks were drawn per channel to tightly enclose bright emission while avoiding negative bowls, which are artifacts typically caused by incomplete uv-coverage. From inspection of the emission across the cube, we masked channels 760~\kms{} to 940~\kms. To mitigate the accumulation of deconvolution errors in the minor cycles, major cycles were triggered frequently (cycleniter = 20). In between major cycles, each channel was inspected and the per channel masks were adjusted to conservatively enclose the remaining emission. In channels 760~\kms{} and 920–940~\kms, the masks were removed after two major cycles, as the masked region was sufficiently indistinguishable from the background fluctuations in the channel. Throughout this process, the CASA log, which reports extract and residual statistics per channel, was also monitored to ensure the extracted emission continually increased in each of the masked channels. After 2-3 major cycles per channel, the extracted emission per channel plateaued; we estimate from comparing the model and residuals from the imaging process that 7$\%$ of the flux density remains in the residual. 

As a final step, we performed a broad, shallow clean across the 800 -- 920~\kms{} channels, which contain the broadest emission in the cube. In these channels, we applied 1\arcmin{}~$\times$~1\arcmin{} masks, centered on the nominal position of Hen 2-10, for the shallow \texttt{tclean}. We increased the frequency of the major cycles (cycleniter = 5) and performed an additional $\sim$20 minor iterations per channel. The \hi{} data product is an image cube, consisting of 107 image channels spaced at 10 \kms{} (47.4 kHz). Table \ref{tab:obs} lists the image parameters for the final data products presented in this work. 

We also reduced the pre-upgrade VLA \hi{} observing programs of Hen 2-10 (projects AK362 and AK343 from \citetalias{Kobulnicky:1995}). VLA project AK362 included observations in the BnA and CnB configurations, where there was $\sim$7 and 2 hours, respectively, of time on source in these observing programs. Project AK343 provided complementary data in the DnC configuration, with $\sim$1 hour time on source. Both AK362 and AK343 were observed with a spectral resolution of approximately 24 kHz (5.1~\kms) and a total bandwidth of 1.56 MHz. The AK343 (DnC) data uniquely included a broader IF, spanning approximately 673–1126~\kms{} ($\sim$3.1 MHz), though at coarser spectral resolution (97 kHz or 20.6~\kms). The higher spectral resolution data sets do not include a comparably broad IF. The archival \hi{} data from VLA projects AK343 and AK362  were reduced following the procedure outlined above. Each observation epoch was reduced independently, and the CASA task \texttt{statwt} was also independently applied. Preliminary imaging was then performed for each epoch to produce dirty images, which were used to identify line-free channels. After excluding edge channels during data reduction, the total velocity coverage was approximately 776–1033~\kms. 

As before, we used the task \texttt{uvcontsub} to subtract the continuum emission. However, the narrow bandwidth of the pre-upgraded VLA observations limited the availability of line-free channels, with no suitable channels at low velocities and only about eight line-free channels above 990~\kms. As a result, when the continuum-subtracted measurement sets from the archival observations were combined for final imaging, residual continuum emission remained at the position of He~2-10. We also tested the alternative continuum subtraction method \texttt{imcontsub}, but it did not yield any improvement.

We combined the higher-resolution archival data with our new observations, using the continuum-subtracted measurement sets from each observing epoch in \texttt{tclean}. However, due to the unreliable continuum subtraction in the archival datasets and their lack of velocity coverage below approximately 776~\kms, the resulting images retained residual continuum emission from the archival data. Therefore, we do not include images or data products from the archival VLA datasets, nor from their combination with the new observations, in the results presented here.

\subsection{CO Data Reduction\label{sec:codatared}}
The CARMA \twco{(1-0)} and \twco{(2-1)} observations were reduced independently in CASA version 4.2. The data were imported using the CASA task \texttt{importmiriad}. Flagging and data inspection were performed prior to standard\footnote{See \url{https://casaguides.nrao.edu/index.php?title=Importing_and_Calibrating_a_Mosaicked_Spectral_Line_Dataset}.} calibrations (e.g., bandpass, complex gain, flux density). 

To improve the sensitivity and image fidelity, we combined the CARMA \twco{(2-1)} data with the archival SMA \twco{(2-1)} (Project Code 2008B-S043), originally presented in \citet{Santangelo_Testi_Gregorini_Leurini_Vanzi_Walmsley_Wilner_2009}. The SMA data were reprocessed following the standard data reduction procedure\footnote{See \url{http://casaguides.nrao.edu/index.php?title=TWHydraBand7_Calibration_3.4}}. Initial imaging was performed with the CASA task \texttt{clean}, with the \textit{imagermode} parameter set to `mosaic', the \textit{mode} set to `velocity', the restoring beam set to `common', the \textit{weighting} set to `briggs' with a robust of 2, and the rest frequency set to 230.538 GHz. The visible emission in each channel was manually masked until the residual image was indistinguishable from noise in a manner similar to that described in Section \ref{sec:vladataredu}. We estimate from comparing the model and residuals from the imaging process that 3$\%$ of the flux density remains in the residual. This is less the uncertainty due to the flux calibration. The final image cubes have 8.3 kHz (10~\kms) wide steps and a 4\farcs4 $\times$ 2\farcs6 beam. We refer to these data as the ``combined \twco{(2-1)}'' data for brevity hereafter. 

For the CARMA \twco{(1-0)} observations, we combined these data with archival ALMA Science Archive (Project Code: 2016.1.00027.S; originally presented in \citealt{Imara:2019}), which had been processed by the standard ALMA pipeline \citep{hunter23}.  In \texttt{tclean}, with the \textit{imagermode} parameter set to `mosaic', the \textit{mode} set to `cube', the restoring beam set to `common', the \textit{weighting} set to `briggs' with a \textit{robust} parameter of 1.5, and the rest frequency set to  115.2712 GHz. The southern CO ``tail'' is very clumpy in the ALMA images shown in \citet{Imara:2019}, visually suggestive of resolved-out emission.  The CARMA data are sensitive to extended emission, but unfortunately, the comparison revealed a problem with the CARMA absolute flux scale that we could not reconcile: the total flux of the CO ``tail'' in a CARMA-only image is only 55\% of that in the ALMA-only image. To test the effect of a flux scale problem in the CARMA data, we multiplied the CARMA amplitude and weights times 2, and the resulting ALMA+CARMA image only differs from the unmodified version by 3\%, so we are confident that our scientific results are not affected by the CARMA flux scale issue, and the spatial and velocity information provide valuable qualitative insights, particularly in relation to the observed \hi{} gas. As an additional check, we imaged the ALMA \twco{(1-0)} data independently, convolving the visibilities to match the resolution of the CARMA \twco{(1-0)} data (4\arcsec{}~$\times$~1\arcsec{}). The smoothed ALMA map shows a similar morphology to the CARMA map, suggesting that both datasets recover comparable spatial distributions. The combined CARMA and ALMA image, created with Briggs weighting robust=1.5, does recover more diffuse emission than the ALMA image presented in \citet{Imara:2019}. In the analysis presented here, we utilize the combined CARMA and ALMA \twco{(1-0)} image cube, which has 3~\kms{} wide steps between 780 -- 900~\kms and a 0\farcs69 $\times$ 0\farcs55 beam; we refer to these data as the combined \twco{(1-0)}.

\subsection{Final Data Products\label{sec:finalprod}}

To determine the noise estimates in the \hi{} and CO data, we sampled a region far from the emission in the line free channels and calculated the standard deviation (STD) of the Median Absolute Deviation (MAD) statistic for all of the extracted regions using the Astropy \citep[version 5.3.4;][]{Astropy:2013,Astropy:2018} function \texttt{mad$\_$std}. This function calculates the MAD statistic multiplied by 1.4826 to approximate the STD of normally distributed data. In these data, the MAD STD is generally 1.2 -- 1.5 $\times$ lower than a traditional rms uncertainty, as the MAD statistic is less sensitive to outliers and large deviations in the dataset. The MAD STD statistics for the \hi{} and CO data are listed in Table \ref{tab:obs} as $\sigma_{HI}$ and $\sigma_{CO}$, respectively.

The reliable measurement of extended emission in interferometric images can be subtle, and has been a research subject for some time. Several techniques were evaluated in \citet{Popping:2012} and revisited for the SoFia package \citep{SoFIA:2014,Serra:2015}, favoring the smooth and clip method of \citet{Serra:2012} for extended emission.  We use this technique to create masked moment images for our \hi{} and CO analyses. Following methods described in \citep{Popping:2012} and \citet{Serra:2012}, we first created a smoothed-and-clipped mask that is the union three components: (1) the original image masked below $\sigma$ times a scalar value A, (2) the image convolved to a beam with twice the original beam minor axis and masked below A$\sigma$ of the convolved image, and (3) the original image with Hanning smoothing in the spectral dimension and then also masked below A$\sigma$ of the Hanning smoothed image. In the \hi{} data, the scalar A was set to 3, and in the \twco{(1-0)} data, it was set to 4.

We then used the CASA task \texttt{immoments} with the \textit{mask} parameter as the smoothed-and-clipped mask to generate the moment 0, 1, 2, and 8 maps and then exported the data into a FITS format for further analysis in Python. To generate the \hi{} column density, N(\hi), map, we converted the smoothed-and-clipped \hi{} moment 0 data to units of K \kms{} and then calculated N(\hi) following Equation (7.155) of \citet{ERA}
\begin{equation}
    \textnormal{N(\hi)} (\textnormal{cm}^{-2}) \approx 1.823 \times 10^{18}  \int \textnormal{T}\; \textnormal{dv},
    \label{eq:column}
\end{equation}
where \(\int\textnormal{T}\; \textnormal{dv}\) is the surface brightness integrated over the line profile in units of K \kms{} and N(\hi) is then in units of cm$^{-2}$. From the \hi{} column density map, we generated the surface density map with the conversion \(\Sigma_{HI}\,[\textnormal{M}_{\odot} \;pc^{-2}] = 8\times10^{-21} \;N_{HI}\,[\textnormal{atoms cm}^{-2}].\)  

\begin{figure}[htb!]
    \centering
    \subfloat{\includegraphics[width=0.48\linewidth]{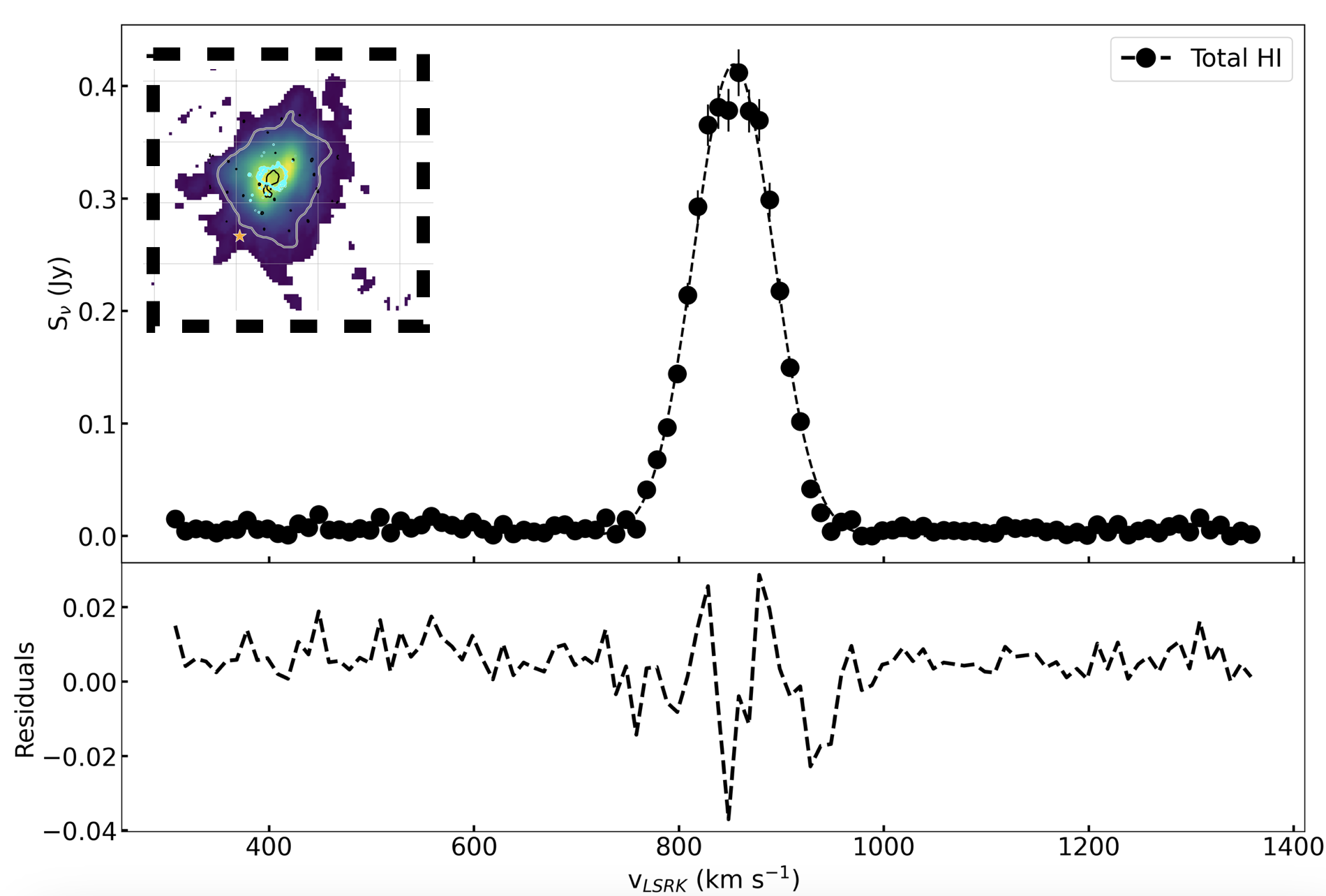}\label{fig:hiline}} \subfloat{\includegraphics[width=0.51\linewidth]{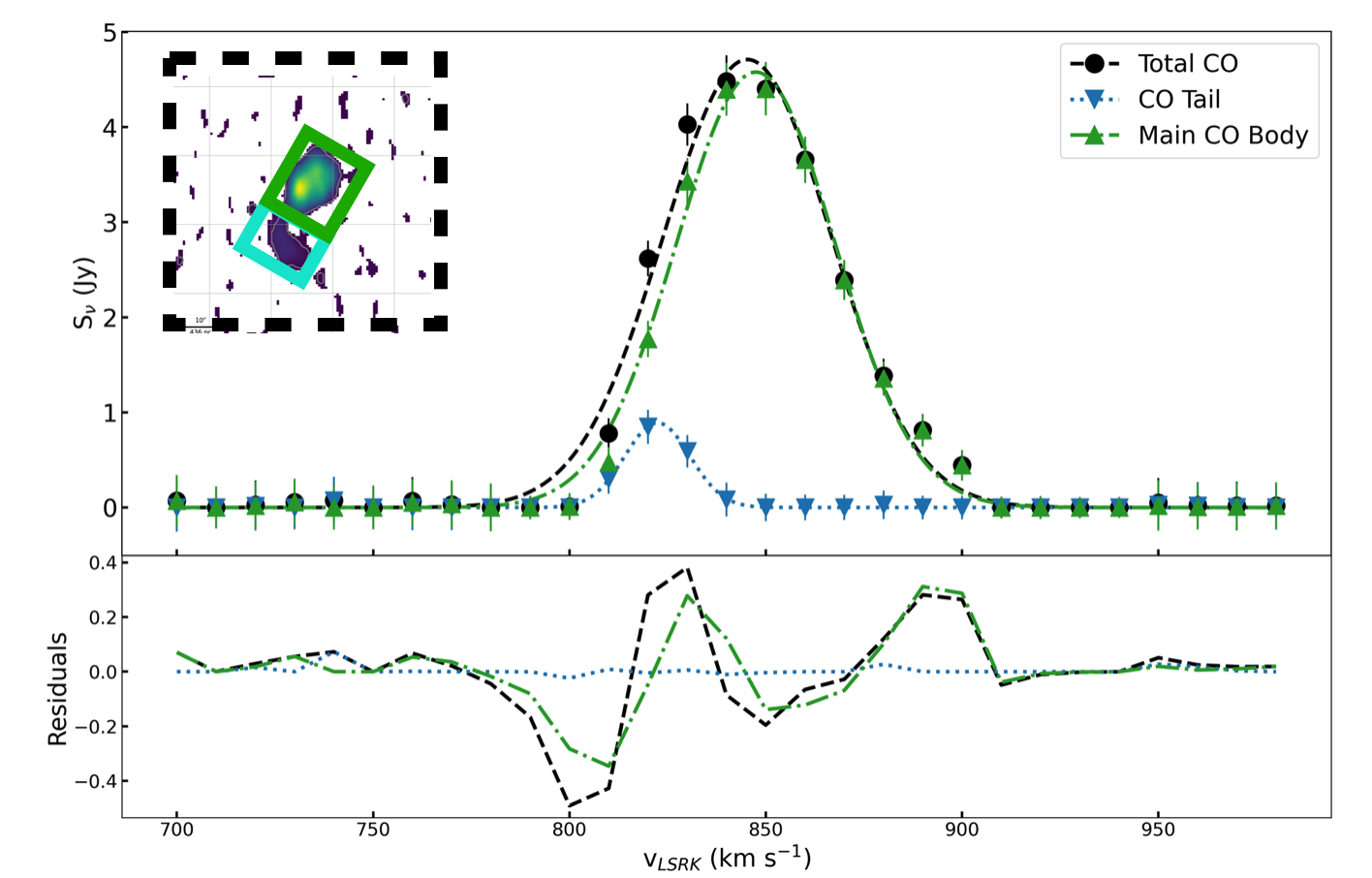}
    \label{fig:coline}}
    \caption{Line profiles in Hen 2-10. \textit{Left, top.} The black symbols represent the surface brightness per velocity channel with 1$\sigma$ uncertainties, and the black dashed curve is the best fit 1D Gaussian. The inset box is the \hi{} integrated intensity map; this FOV is $\sim$4\farcm5 $\times$ 4\farcm5. \textit{Left, bottom.} The black dashed curve shows the residuals ( = measurement - model). \textit{Right, top.} The \twco{(2-1)} line profiles for three extracted regions are shown, and the inset box is the \twco{(2-1)} integrated intensity map with the extracted regions overlaid. The region of main body of CO emission is marked with a green rectangle; its line profile is the green dash-dotted curve. The CO ``tail'' is the cyan rectangle; its line profile is the blue dotted curve. The black dashed curve is the emission within the black dashed rectangle, representing the total CO mass. The \twco{(1-0)} profiles are qualitatively consistent with the profiles shown here. \textit{Right, bottom.} The black dashed, green dash-dotted, and dotted blue curves show the residuals for the total, the main CO body, and the CO ``tail'', respectively. See Table \ref{tab:regions} for qualitative details of the regions.}
    \label{fig:line}
\end{figure}

We defined spatial masks to generate line profiles and to estimate the masses of different regions in the spectral cubes. For the \hi{} data, we set a 4\arcminn{} $\times$ 4\arcminn{} rectangle to enclose the majority of the 3$\sigma$ emission. In the \twco{(2-1)} data, we applied three spatial masks: one capturing the main body of CO emission, a second for the CO ``tail'', and a final mask encompassing the total CO emission. Figure \ref{fig:line} provides a visual reference, and the parameters of the extraction apertures are in Table \ref{tab:regions}. For each region, we extracted the intensity defined within the spatial mask for each velocity channel with the \texttt{RectangularAperture} function from the Python package Photutils \citep[version 1.11.0;][]{Bradley:1.11}. We integrated the extracted intensity, in units of Jy beam$^{-1}$, over the solid angle subtended by the region per velocity bin, yielding the integrated flux density in units of Jy. To estimate the mass, the integrated intensity defined within the spatial mask from the moment 0 map was integrated over the solid angle subtended by a region, yielding the integrated flux density in units of Jy~\kms. The uncertainties in the integrated flux density measurement are given by \[\sigma = \sigma_{chan}\;\Delta V \sqrt{N_{chan}},\]
where $\sigma_{chan}$ is the quadrature of the MAD STD statistic for a channel and the flux calibration uncertainty, $\Delta V$ is the width of the channel in \kms, and $N_{chan}$ is the number of channels in the integration. For VLA 21-cm observations, the flux calibration uncertainty is typically 3\%\footnote{\url{https://science.nrao.edu/facilities/vla/docs/manuals/oss/performance/fdscale}}; for ALMA Band 3 (115 GHz), it is $\sim$10\% \citep{Fomalont:2014}. The uncertainty in the integrated flux density is propagated forward for the mass calculations.

We find the total \hi{} (M$_{HI}$) and molecular (M$_{H2}$) masses from the integrated flux density. Following \citet{Roberts1975} and Equation (4) of \citetalias{Kobulnicky:1995}, the \hi{} mass
not accounting for Helium is
\begin{equation}
    M_{HI} (M_{\odot}) = 2.365 \times 10^{5}\; \textnormal{D}^2 \; \int \textnormal{S}_{HI} \; \textnormal{dv},
    \label{eq:himass}
\end{equation}
where D is the distance in Mpc, \(\int\textnormal{S}_{HI}\; \textnormal{dv}\) is the integrated \hi{} flux in units of Jy \kms, and M$_{HI}$ is then in units of \Msun. From Equation (3) of \citet{Bolatto:2013}, the total molecular mass is
\begin{equation}
    M_{H2} = 1.05 \times 10^4 \left(\frac{X_{CO}}{2 \times 10^{20}}\right) \frac{\textnormal{D}^2}{(1+z)} \int \textnormal{S}_{CO_{1\rightarrow0}} \; \textnormal{dv}
    \label{eq:mol}
\end{equation}
where X$_{CO}$ is the conversion factor from the CO intensity to H$_2$ column density in units of cm$^{-2}$ (K \kms)$^{-1}$, D is the distance in Mpc, $z$ is the redshift, and \(\int\textnormal{S}_{CO_{1\rightarrow0}}\; \textnormal{dv}\) is the integrated \twco{(1-0)} flux in units of Jy~\kms. Estimating the total molecular mass depends on the CO-to-$H_2$ conversion factor ($\text{X}_{CO}$), with a typical value of 2 $\times$ 10$^{20}$ cm$^{-2}$ (K \kms)$^{-1}$ \citep[e.g.,][]{Bolatto:2013}. We use 3 $\times$ 10$^{20}$ cm$^{-2}$ (K \kms)$^{-1}$ for consistency with \citetalias{Kobulnicky:1995} and \citet{Santangelo_Testi_Gregorini_Leurini_Vanzi_Walmsley_Wilner_2009}; we also propagate an X$_{CO}$ uncertainty factor of 30\% throughout the calculations \citep[][]{Bolatto:2013}. We estimate M$_{H2}$ with the integrated flux from the \twco{(1-0)} transition as well as \twco{(2-1)} transition. For the latter, the \twco{(2-1)} transition must be scaled as an estimate of the \twco{(1-0)} line using a line ratio, R$_{21}$,
\begin{equation}
    S_{CO_{1\rightarrow0}} \approx \frac{S_{CO_{2\rightarrow1}}}{R_{21}} \; \left(\frac{\nu_{115 \;GHz}}{\nu_{230 \;GHz}}\right)^{2}.
\end{equation}
From single dish observations, \citet{Baas:1994} estimates R$_{21}$ $\sim$ 0.97, but \citet{Leroy:2013} find that while starbust galaxies have enhanced line ratios at galaxy centers, generally, line ratios are anticorrelated with distance from the center. The value of R$_{21}$ for individual regions within Hen 2-10 may be inconsistent with a global value, which reflects environmental dependencies \citep[e.g.,][]{Leroy:2022}. As the R$_{21}$ is uncertain for Hen 2-10, we adopt the value of \citet{Baas:1994} and propagate a 10\% uncertainty.  

With the Python package Pvextractor\footnote{\url{https://pvextractor.readthedocs.io/en/latest/}} version 0.3 and Spectral-Cube \citep[version 0.6.5;][]{Ginsburg2019}, we generated position-velocity (p-v) slices from the \hi{} and \twco{(1-0)} data, where the position angle (PA) that defines the slice is measured from the north (+y-axis) to east (--x-axis) in the image plane, for consistency with previous works (e.g., \citetalias{Kobulnicky:1995}; \citealt{Imara:2019}). Note, this is different than the definition implemented in Python and Pvextractor, which use the +x-axis to define 0\ddeg. For the \hi{} data, we inspected p-v slices between PA = 0\ddeg{} and 180\ddeg{} in increments of 5\ddeg, with each slice centered at the nominal position of Region A.  We calculated the intensity weighted mean in each 10 \kms{} channel across a p-v slice and performed a linear fit to the intensity weighted mean per channel to identify the p-v slice with the steepest gradient.  In the \hi{} data, we extracted three final slices for further analysis: one to sample H$\alpha$ outflows at PA = 49\ddeg{} \citep{Mendez:1999}, one for the major axis of rotation identified by \citetalias{Kobulnicky:1995} at PA = 130\ddeg, and one for the steepest gradient identified in this analysis at PA = 100\ddeg. For these p-v slices, the sampling along the slice was set to 1 pixel, the width of the path was set equal to the minor axis of the synthesized beam (13\arcsec), and the slice is anchored on Region A. In the \twco{(1-0)} data, we similarly extracted slices at PA = 49\ddeg{} and 130\ddeg, both anchored at Region A. We then shifted the anchor point to be the midpoint between Regions A and B in order to intersect the `V-shaped' emission in the north and the southern ``tail''. This final slice has a PA = 180\ddeg. In the \twco{(1-0)} data, the width of the path is 5 times the minor axis of the synthesized beam (0.55\arcsec).

\section{Results}\label{sec:results}

\begin{deluxetable*}{cccccccccc}[htb!]
\tablewidth{0.8\textwidth} 
\tablecaption{Region Properties and Results \label{tab:regions}}
\tablehead{
\colhead{Region} & \colhead{R.A.}  & \colhead{Dec.} & \colhead{Line Center} & \colhead{Line Width} & \multicolumn{3}{c}{Aperture} & \colhead{S$_{\nu}$} &\colhead{M}\\ \cmidrule{6-8}
\colhead{} & \colhead{(J2000)} & \colhead{(J2000)} & \colhead{(LSRK)} & \colhead{(FWHM)} & \colhead{Width} & \colhead{Height} & \colhead{PA} & &\colhead{$\times$ 10$^7$} \\
\colhead{} & \colhead{(HH:MM:SS)} & \colhead{(DD:MM:SS)} & \colhead{(\kms)} & \colhead{(\kms)} & \colhead{(\arcsecc)} & \colhead{(\arcsecc)} & \colhead{(\ddeg)} &\colhead({Jy~\kms)}  &\colhead{(\Msun)}
}
\startdata
\hi{} Total & 08:36:14.65 & --26:24:26.41 & 858 & 117 & 240 & 240 & 0 & 8.8 $\pm$ 0.3 & 16.9 $\pm$ 0.5\\
\twco{(1-0)} Total & 08:36:15.52 & --26:24:36.36 & 849 & 58 & 42 & 46 & 0 & 111 $\pm$ 11 & 14.1 $\pm$ 4.4\\
\twco{(1-0)} Body & 08:36:15.18 & --26:24:36.41 & 850 & 58 & 15 & 17 & --30 & 88 $\pm$ 9 &11.1 $\pm$ 3.5\\
\twco{(1-0)} Tail & 08:36:15.70 & --26:24:49.00 & 826 & 30 & 15 & 11 & --30 & 20 $\pm$ 2 &2.5 $\pm$ 0.8\\
\twco{(2-1)} Total & 08:36:15.52 & --26:24:36.36 & 845 & 50 & 42 & 46 & 0 & 268 $\pm$ 27 & 8.8 $\pm$ 2.9\\
\twco{(2-1)} Body & 08:36:15.18 & --26:24:36.41 & 847 & 48 & 15 & 17 & --30 & 239 $\pm$ 24 & 7.8 $\pm$ 2.6\\
\twco{(2-1)} Tail & 08:36:15.70 & --26:24:49.00 & 822 & 20 & 15 & 11 & --30 & 25 $\pm$ 27 & 0.8 $\pm$ 0.3\\
\enddata
\tablecomments{Columns 2 -- 3 list the nominal center coordinates of presented regions; as all are unresolved, the listed coordinates are for general reference. Columns 4 and 5 give the modeled line centers and line widths for each transition and region. Columns 6 -- 8 specify the inputs to the Photutils function \texttt{RectangularAperture}. Column 9 is the measured integrated flux density of the region, and Column 10 is the calculated \hi{} mass in the first row and otherwise is the molecular mass. The M$_{H2}$ calculated for \twco{(2-1)} uses R$_{21}$ = 0.97.}
\end{deluxetable*}

\subsection{Distribution of HI Gas\label{sec:distroHI}}
\begin{figure*}[htb!]
    \centering
    \subfloat{\includegraphics[width=0.42\linewidth]{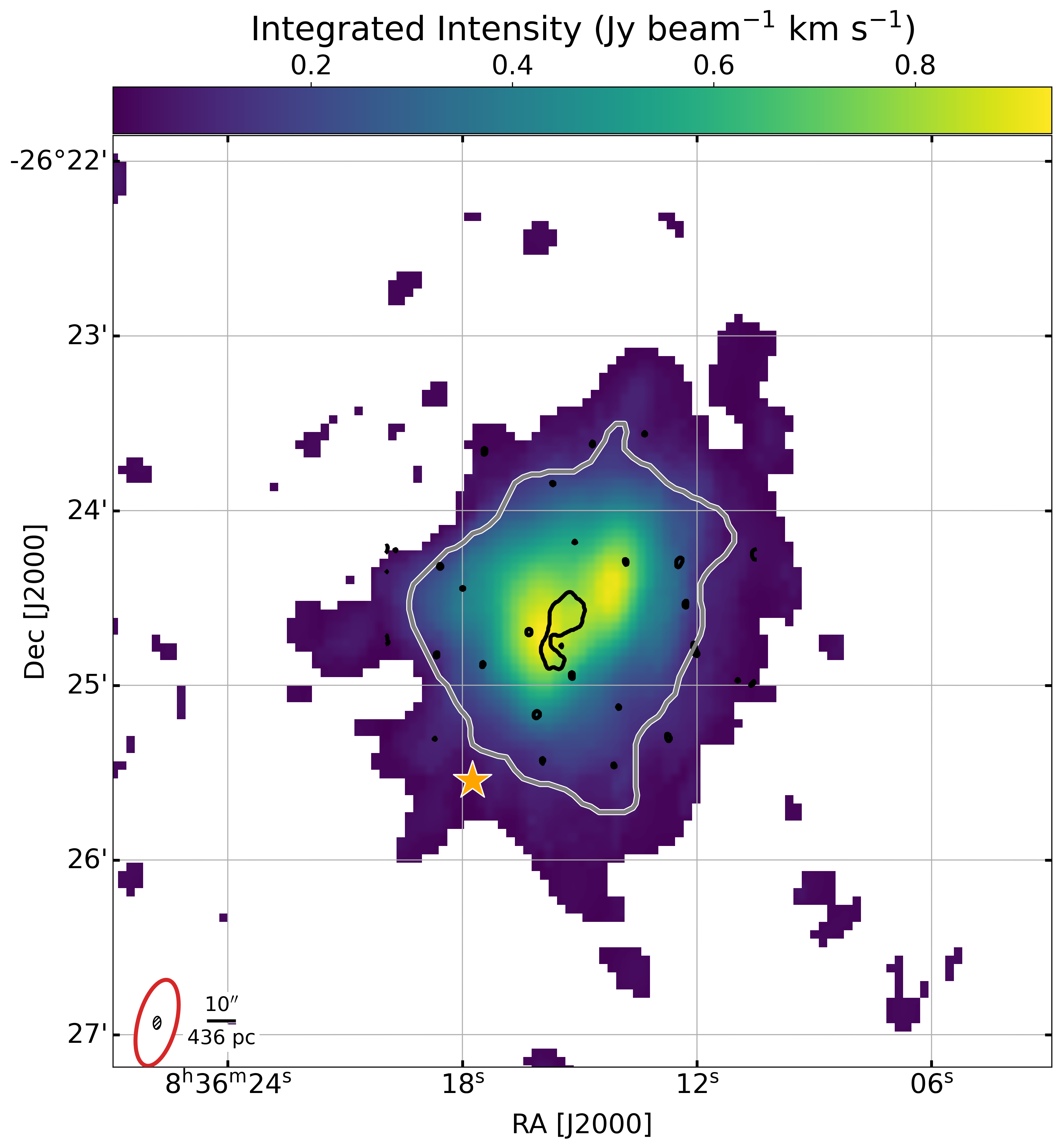}\label{fig:HImom0}}
    \subfloat{\includegraphics[width=0.42\linewidth]{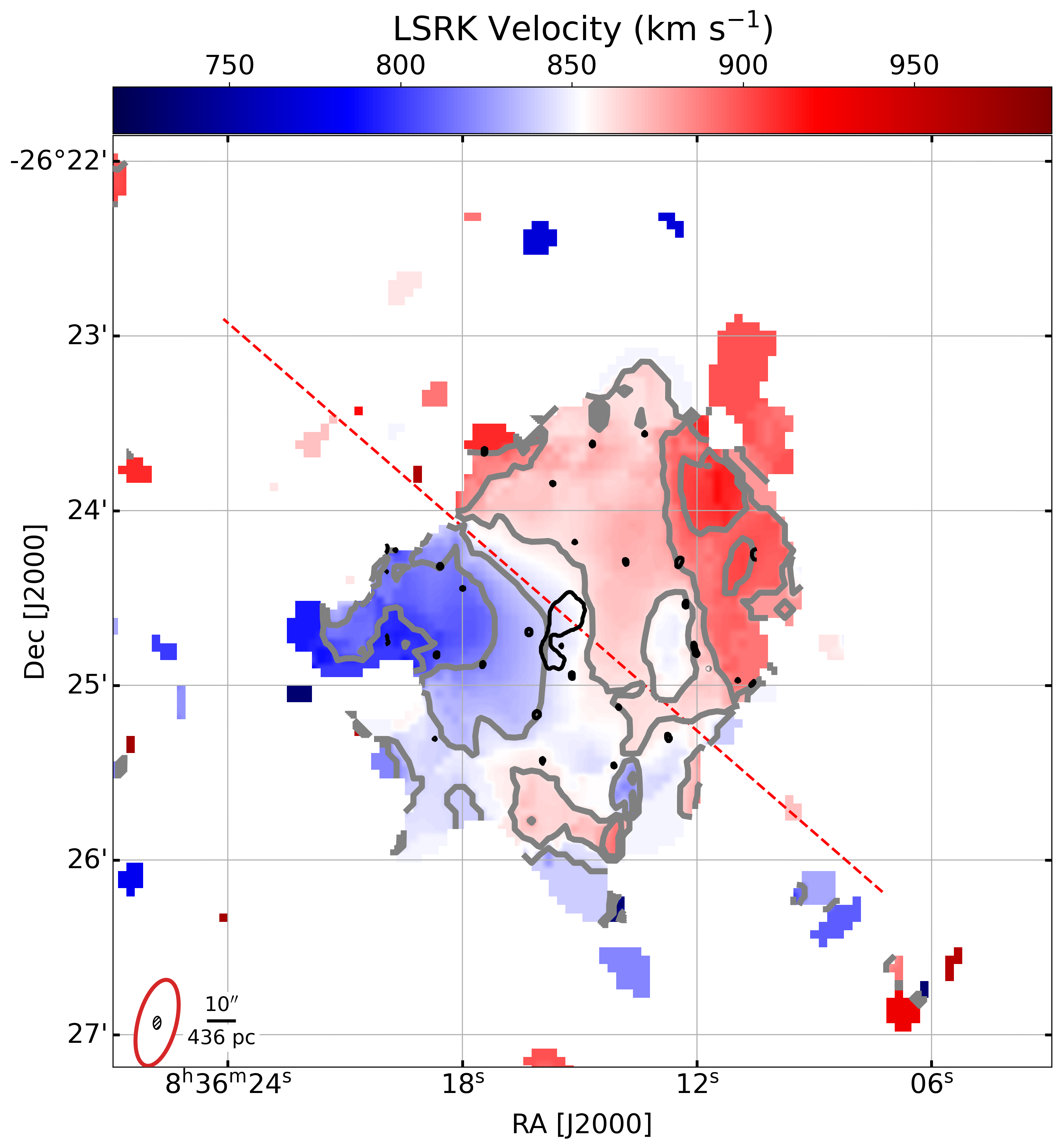}
    \label{fig:HImom1}}
    \quad
    \subfloat{\includegraphics[width=0.42\linewidth]{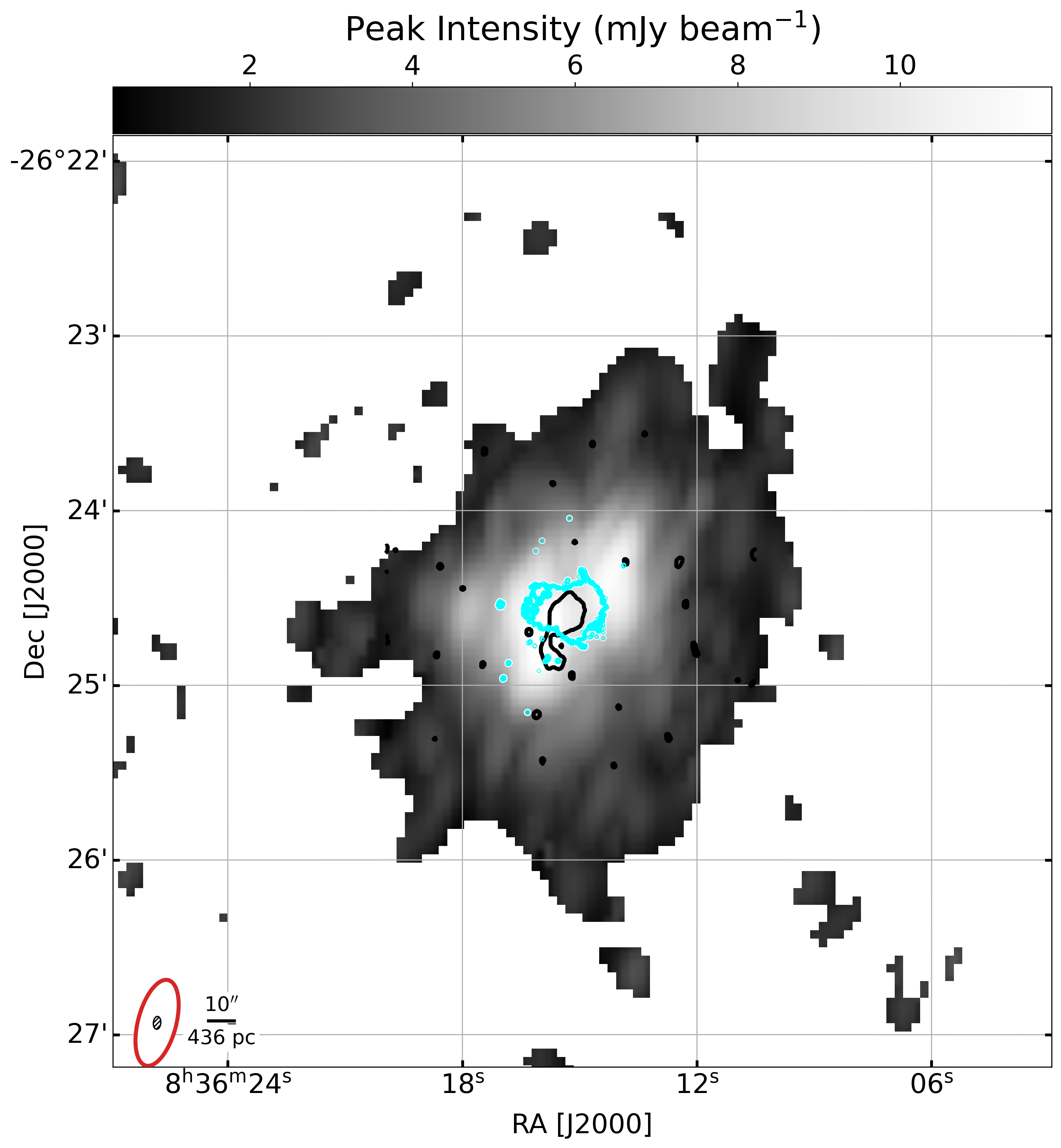}
    \label{fig:HImom8}}
    \subfloat{\includegraphics[width=0.42\linewidth]{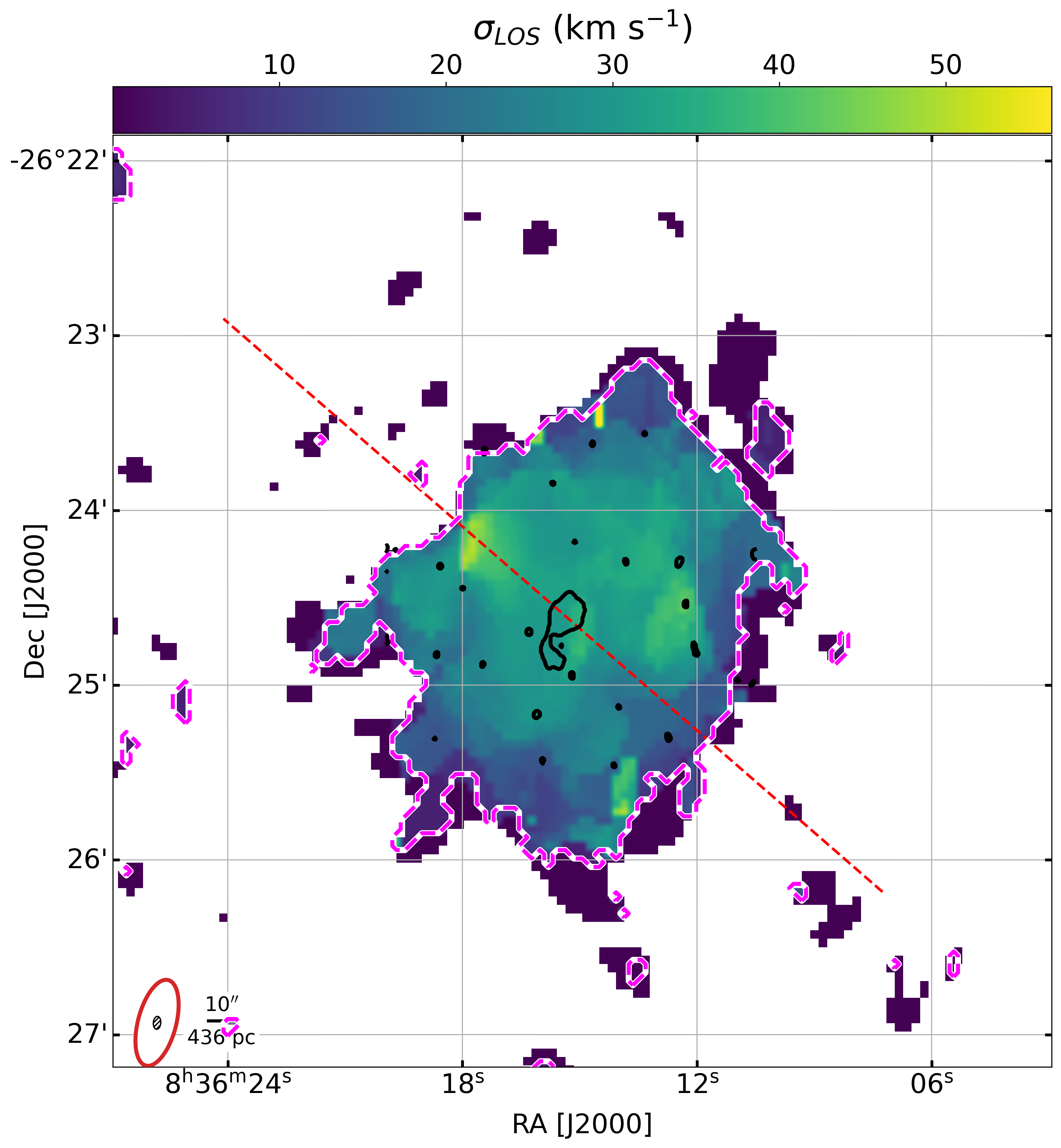}
    \label{fig:HImom2}}
    \caption{\hi{} maps of (a) the integrated intensity (b) intensity weighted velocity field, (c) the peak intensity, and (d) the intensity weighted velocity dispersion, integrated over 730 -- 990~\kms. In each panel, the black contour shows 5$\sigma_{CO(2-1)}$. The synthesized \hi{} beam is the red ellipse in the lower left, and the combined \twco{(2-1)} beam is overlaid as the black hatched ellipse at the center of the \hi{} beam. In all panels, the raster is masked below 3$\sigma_{HI}$. In (a), the gray contour is 6$\sigma_{HI}$, and the orange star represents approximately the tip of \hi{} tail seen in  \citetalias{Kobulnicky:1995}. In (b), isovelocity contours are shown in increments of 20~\kms. In (c) the cyan contour is the H$\alpha$ ``footprint''. In (d), the red dashed line intersects both the NE and SW bubbles from Figure \ref{fig:f555FinderA}, and the dashed fuchsia contour is at the \hi{} surface density of 1 \Msun{} pc$^{-2}$. The two regions of high velocity dispersion are denoted with the orange arrows (see text for discussion). Note, the $\sigma_v$ $\sim$ 60~\kms{} clump near (RA, Dec.) = (08:36:14.5, --26:23:28) is likely an artifact in the image.
    }
    \label{fig:HImom}
\end{figure*}

In Figure \ref{fig:HImom0}, the \hi{} integrated intensity map reveals an extended atomic envelope reaching more than 1\arcminn{} ($> 2.5$~kpc) from the nominal center of Hen 2-10. The global \hi{} line profile (Figure \ref{fig:hiline}) is consistent with earlier measurements and shows hints of a two-component structure, which is associated with two prominent peaks in the \hi{} emission. However, these peaks are sufficiently broadened and blended, both by the coarse beam and their intrinsic smoothness, that isolating them kinematically offers limited additional insight in these data. The two \hi{} peaks themselves are offset from Regions A and B, as previously noted by \citetalias{Kobulnicky:1995}. Though there is \hi{} emission located at the nominal position of the \citetalias{Kobulnicky:1995} \hi{} tail (see orange star in Figure \ref{fig:HImom0}), the emission is not more significant, or striking, than other features seen at the edge of the 3 or 5$\sigma$ emission. The \citetalias{Kobulnicky:1995} data included D-, C-, and B-array observations, but their \hi{} maps (e.g. their Figure 6) do not indicate additional extended \hi{} features beyond their \hi{} tail. Thus, we cannot confirm that there is a significant \hi{} tail in the SE direction. 

\begin{figure}[htb!]
    \centering
    \includegraphics[width=0.8\linewidth]{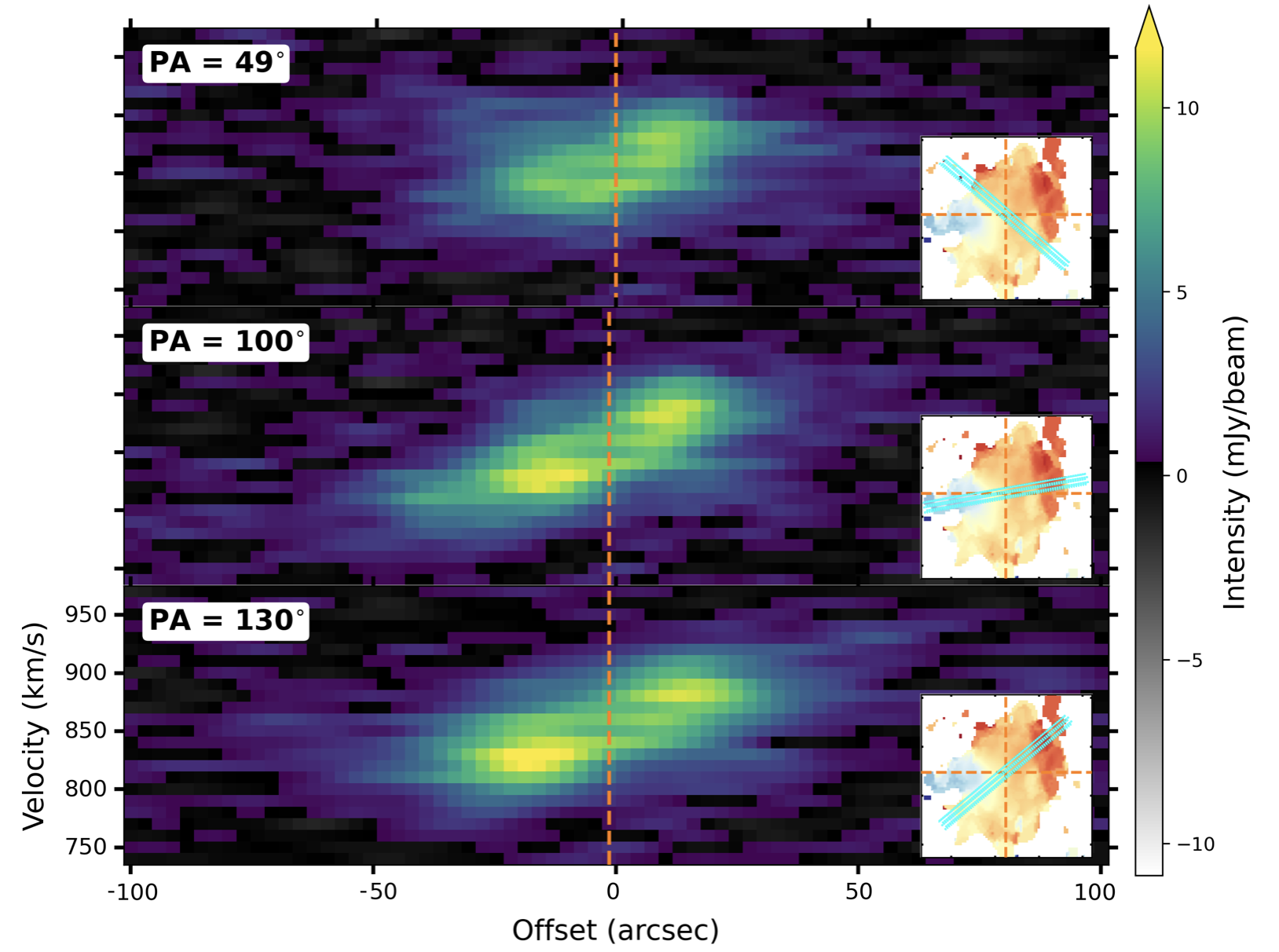}
    \caption{Position-velocity diagrams for the \hi{} data. The inset raster map is the \hi{} moment 1 data, marking the paths of the slices.  All slices are anchored on Region A, which is marked with two dashed orange lines in the raster map and the vertical dashed orange line on the p-v slice. Emission in the position–velocity diagram is displayed with a two-tone colormap: signal above $\sigma_{HI}$ appears in color, and fainter emission is shown in grayscale.}
    \label{fig:pv_hi}
\end{figure}

In the intensity weighted velocity field map (Figure \ref{fig:HImom1}), the isovelocity contours are asymmetric but exhibit a spider-like shape, suggesting that there is a rotating \hi{} disk.  \citetalias{Kobulnicky:1995} reported the major axis of rotation is at PA = 130\ddeg{} (as measured from the +y axis), which is nearly perpendicular to the PA of the H$\alpha$ outflows (PA = 49\ddeg; \citealt{Mendez:1999}). The major and minor axes of rotation typically have the steepest and shallowest gradients, respectively, and while the 49\ddeg{} slice is arguably the shallowest gradient, the 100\ddeg{} slice is nearly as steep as the 130\ddeg{} slice (see Figure \ref{fig:pv_hi}). Furthermore, the 100\ddeg{} slice intersects the most blueshifted material in the east, suggesting that the disk may be warped or disturbed kinematically. In the channel maps (Figure \ref{fig:HIchan}), this emission is first seen in the $\sim$ 770~\kms{} channel. Further indications of disturbed kinematics are in the southwestern part of Hen 2-10 in the Figure \ref{fig:HImom1}, which has alternating regions of red- and blueshifted material, similar to dispersion dominated envelopes. 

In the intensity weighted velocity dispersion map (Figure \ref{fig:HImom2}), the bulk of the neutral atomic gas has dispersions between 10 -- 30~\kms, with a mean of $\sim$8~\kms, which is typical of the \hi{} in dwarf galaxies on $\lesssim$kpc scales \citep{Stil:2002}. However, there are two beam-sized regions with higher velocity dispersions of $\sim$50~\kms; these are offset by roughly 35\arcsec{} (1.5 kpc) to the northeast and south of the starbursting regions. Notably, the northeastern region lies along the position angle of the NE outflow ($\sim$49\ddeg). Usually, \hi{} velocity dispersion is anticorrelated with distance from the galactic center \citep[e.g.,][]{Stil:2002,Tamburro:2009,Eibensteiner:2023}, so some mechanism (e.g., supernovae) may be driving the higher values seen in these regions.

\subsubsection{HI Mass}
\begin{figure}[htb!]
    \centering
    \subfloat{\includegraphics[width=0.4\linewidth]{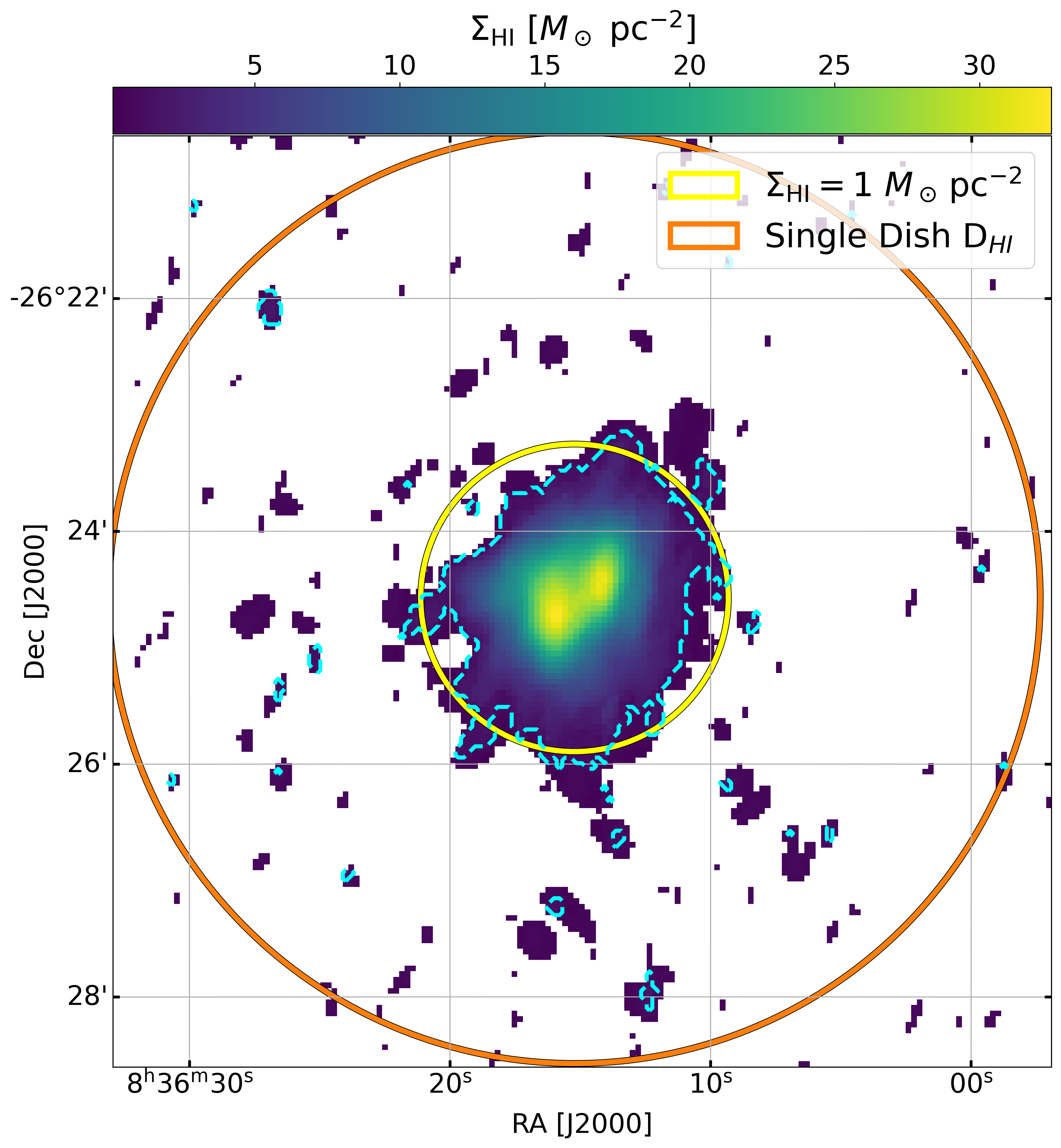}\label{fig:surfaced}}
    \subfloat{\includegraphics[width=0.5\linewidth]{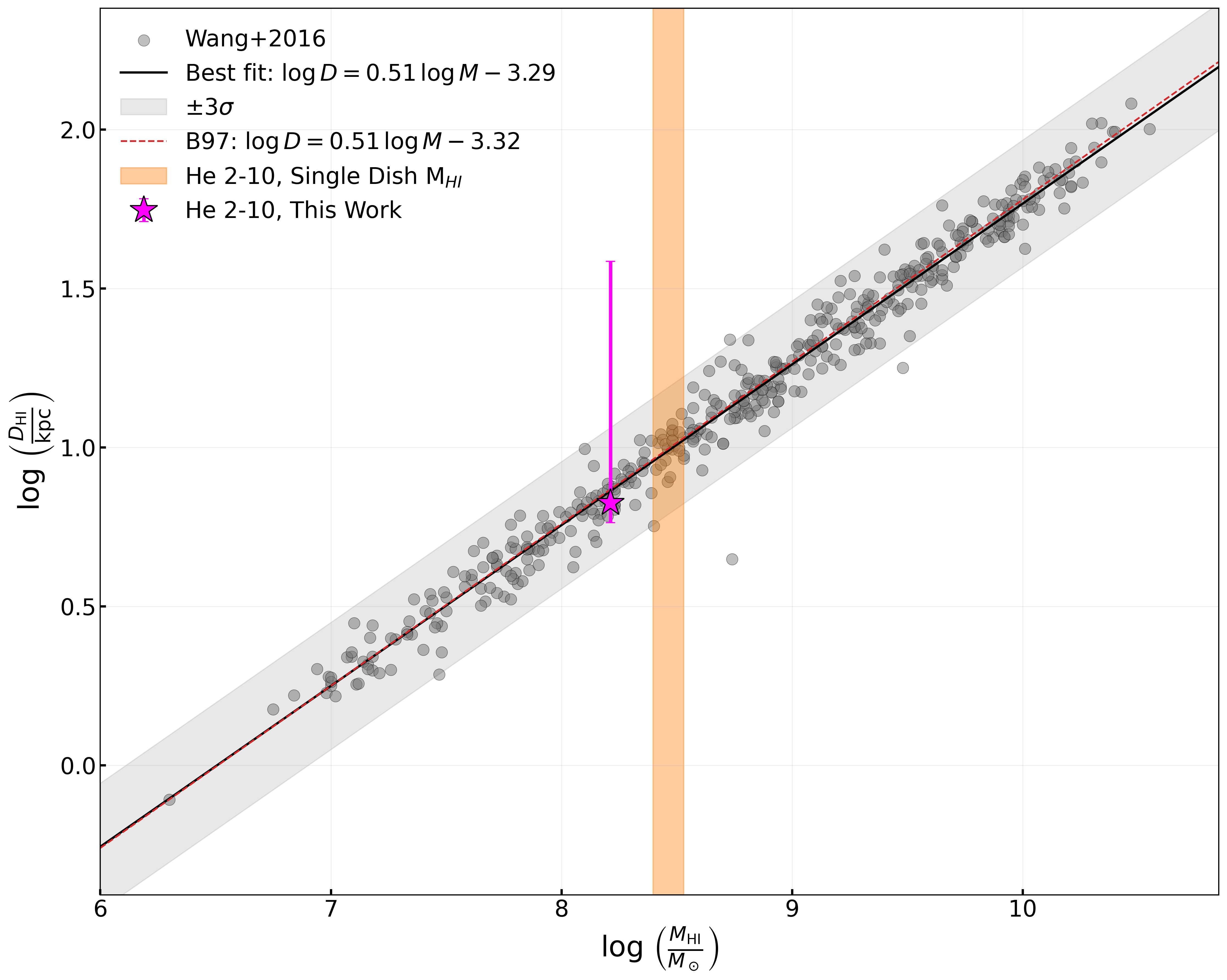}\label{fig:dhi}}
    \caption{\textit{Left:} Raster map of $\Sigma_{HI}$, masked below 3$\sigma_{HI}$. The diameter of the yellow circle is D$_{HI}$; the dashed cyan contour traces $\Sigma_{HI}$ = 1 \Msun{} pc$^{-2}$. The diameter of the orange circle is 4\arcminn, which is D$_{HI}$ estimated from single dish \hi{} mass estimates. \textit{Right:} Reproduction of Figure 1 of \citet{Wang:2016}, where the gray circles are galaxies in their sample, the red dashed line is the relationship from \citet{Broeils:1997}, and the solid black line is Equation \ref{eq:wang2016}, with the shaded gray region marking the $\pm$3$\sigma$. The fuchsia star marks the mass enclosed by a circle with a diameter of 7 kpc to represent Hen 2-10. We do not correct D$_{HI}$ for inclination; the uncertainties represent the range in D$_{HI}$ for inclination angles spanning 15--80\ddeg. The orange, vertical shaded region marks the range of \hi{} masses derived from single dish observations.}
    \label{fig:dvsm}
\end{figure}

\citet{Broeils:1997} identified a strong empirical link between a galaxy’s \hi{} mass and its \hi{} diameter, defined at a surface density of $\Sigma_{\mathrm{HI}} = 1$ \Msun{} pc$^{-2}$. This $D_{\mathrm{HI}}$–$M_{\mathrm{HI}}$ relationship also holds for dwarf galaxies \citep{Swaters:2002,Begum:2008}, so therefore we can leverage this empirical relationship to infer the diameter of the \hi{} emission in the absence of observations at those spatial scales. Using a large literature sample that includes spirals, early-type systems, and dwarfs, \citet{Wang:2016} confirmed the tight correlation and expressed it as
\begin{equation}
    \log(D_{HI}) = 0.506\log(M_{HI})-3.291,
    \label{eq:wang2016}
\end{equation}
where $D_{\mathrm{HI}}$ is in kpc and $M_{\mathrm{HI}}$ in \Msun. Following \citet{Wang:2016}, we define $D_{\mathrm{HI}}$ as two times the radius at which the azimuthally averaged surface density drops to 1 \Msun{} pc$^{-2}$ by measuring $\Sigma_{\mathrm{HI}}$ in concentric annuli about the optical center. We obtain $D_{\mathrm{HI}}\simeq$ 2.55\arcminn{} or 6.69 kpc at a distance of 9 Mpc. Figure \ref{fig:surfaced} overlays this diameter on the \hi{} surface-density map of Hen 2-10. Integrating within a circle of this size yields an \hi{} mass of $(16.3 \pm 0.5)\times10^{7}$ \Msun{} from Equation \ref{eq:himass}. Comparing the \hi{} mass enclosed by D$_{HI}$, Hen 2-10 is within the 3$\sigma$ scatter of Equation \ref{eq:wang2016}, which is shown in Figure \ref{fig:dhi}. \citet{Nguyen2014} estimate the inclination to be $\sim$39\ddeg; an $i>$60\ddeg{} would place Hen 2-10 outside the $+3\sigma$ scatter. 

The morphology of the \hi{} emission at the edge of the moment maps suggest the presence of more diffuse, faint \hi{} emission has not been recovered in these data. The ``smooth and clip'' method used to generate the \hi{} maps further supports such an extended \hi{} envelope, which appears in the 3-component mask as well as masks created from only 1 of those components (Section \ref{sec:finalprod}). The column density sensitivity of the data is approximately $2.5 \times 10^{19}$ cm$^{-2}$ (1$\sigma$), so any extended structure below this threshold would remain undetected in these data. The \hi{} mass measured in these interferometric observations represents only 40–60\% of the single-dish values from \citet{Allen1976}, \citet{Sauvage1997}, and \citet{Meyer:2004}. Taking the range of \hi{} mass from single dish estimates ($2.5-3.4\times 10^8$ \Msun), the diameters are between 9.0 -- 10.5 kpc. At the top of this range, the \hi{} envelope would have a diameter of $\sim$4\arcmin{} (see Figure \ref{fig:surfaced}), which is the extent of the region we extracted \textit{a priori} in Section \ref{sec:analysis}. Integrating over a 4\arcmin{}$~\times~$4\arcmin{} region gives $(16.8 \pm 0.5)\times10^{7}$ \Msun, which is consistent with the mass estimated enclosed by D$_{HI}$ within uncertainties; both are consistent with the mass reported by \citetalias{Kobulnicky:1995}. If Hen 2-10 follows the $D_{\mathrm{HI}}$–$M_{\mathrm{HI}}$ relationship, then the measured single dish mass implies an extended \hi{} reservoir, unless the inclination angle is quite low.

\subsection{Distribution for CO Gas\label{sec:distCO}}

\begin{figure*}[htb!]
    \centering
    \subfloat{\includegraphics[width=0.42\linewidth]{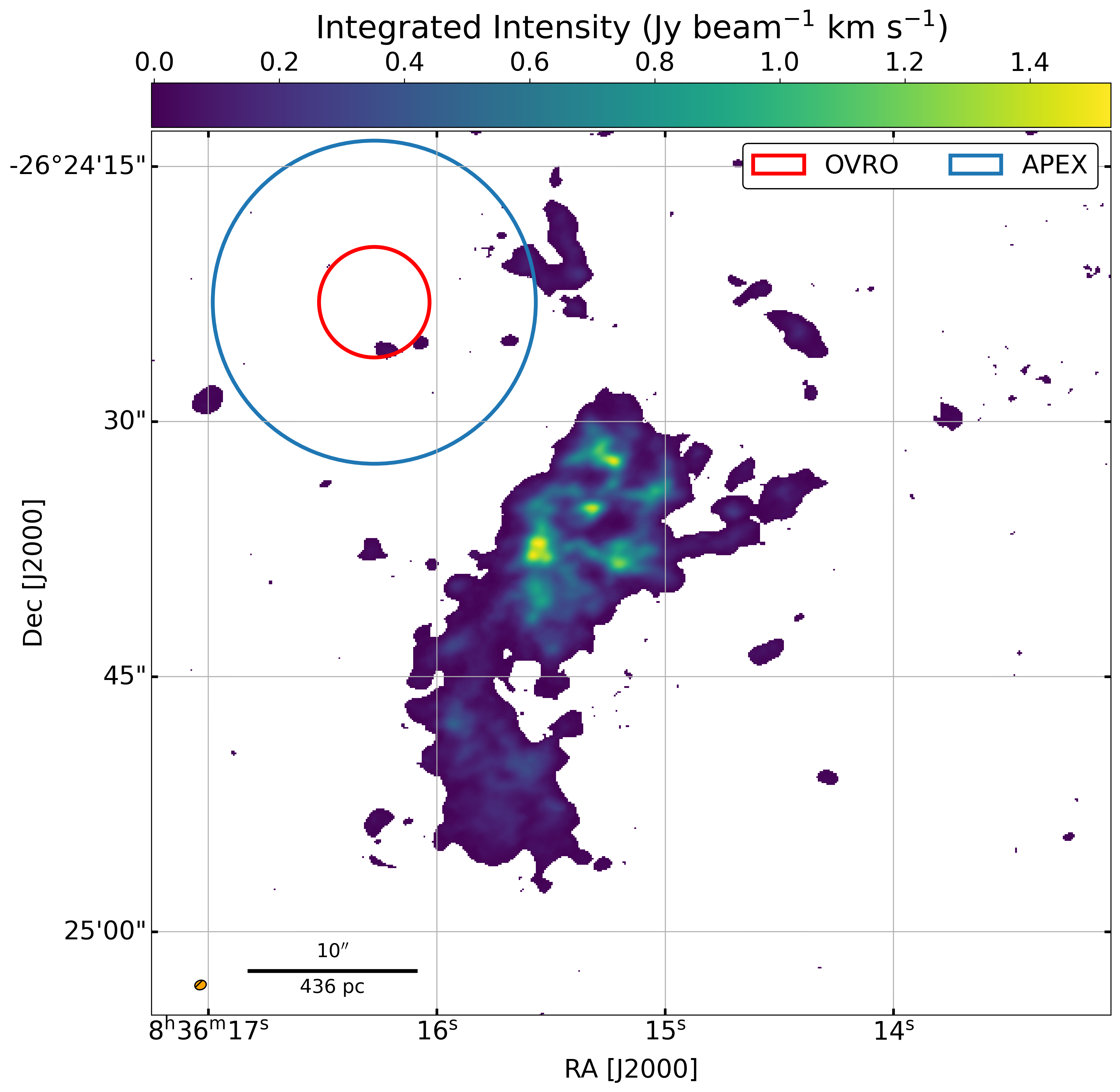}\label{fig:10mom0}}
    \subfloat{\includegraphics[width=0.42\linewidth]{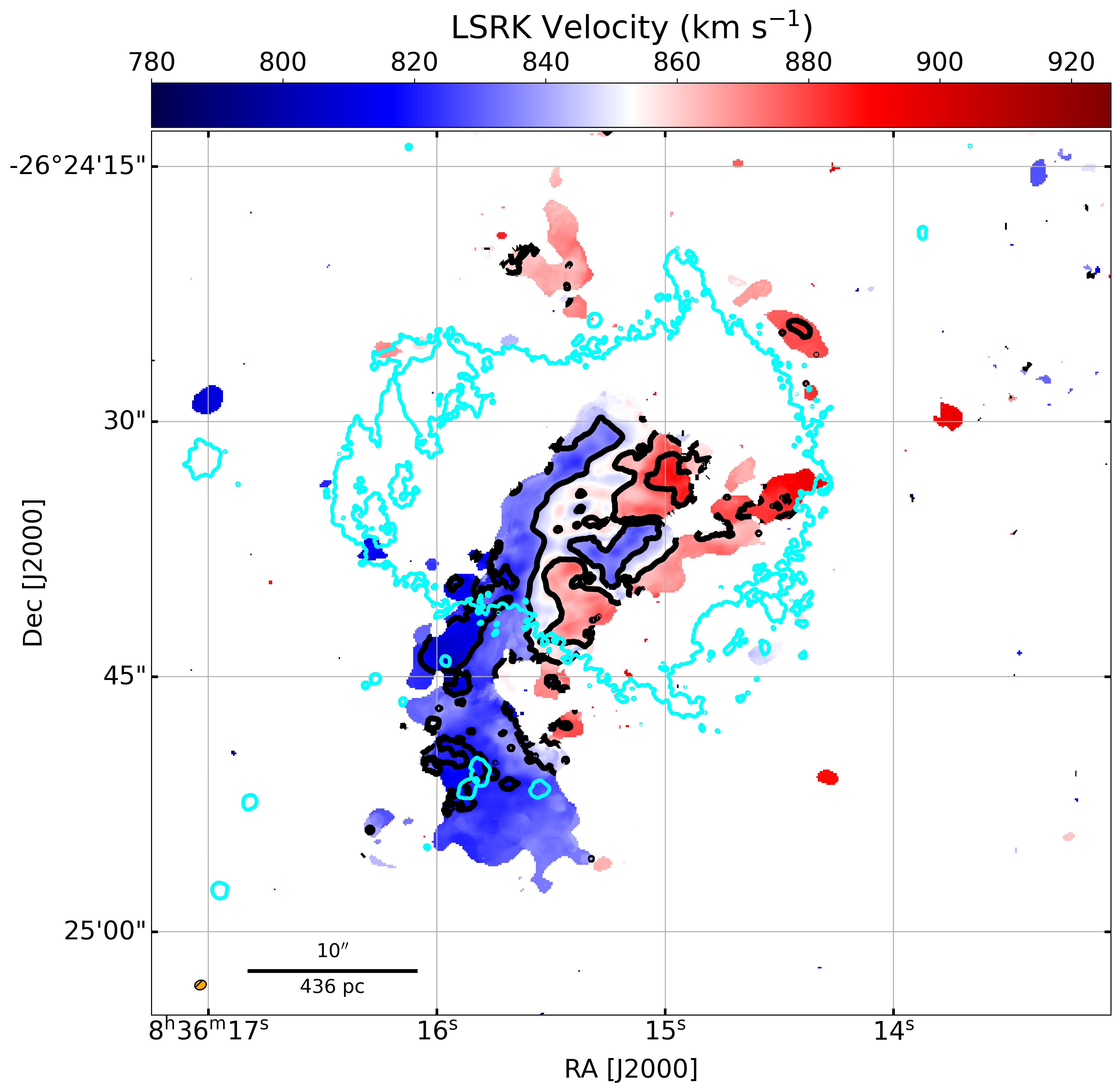}
    \label{fig:10mom1}}
    \quad
    \subfloat{\includegraphics[width=0.42\linewidth]{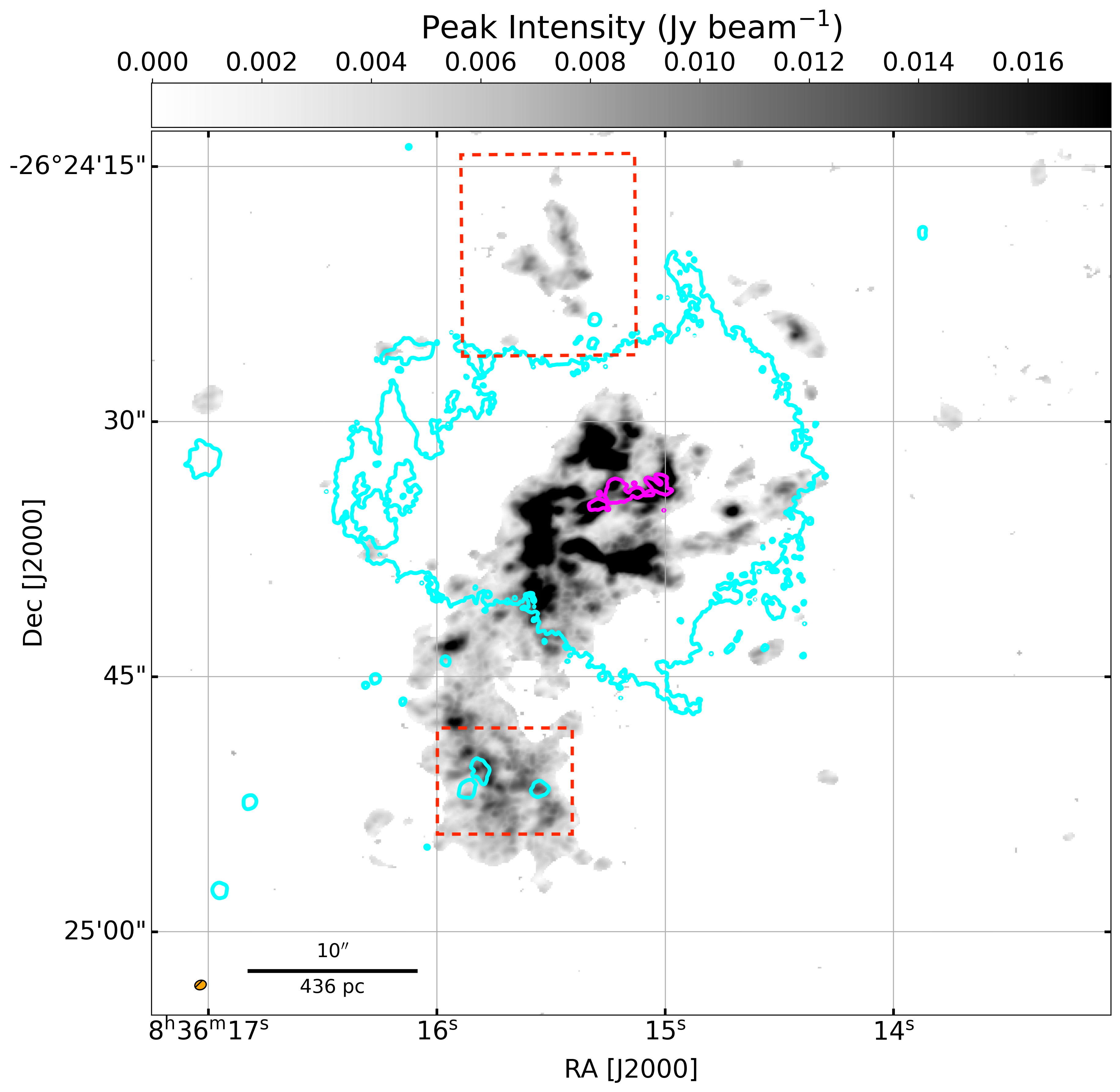}\label{fig:10mom8}}
    \subfloat{\includegraphics[width=0.42\linewidth]{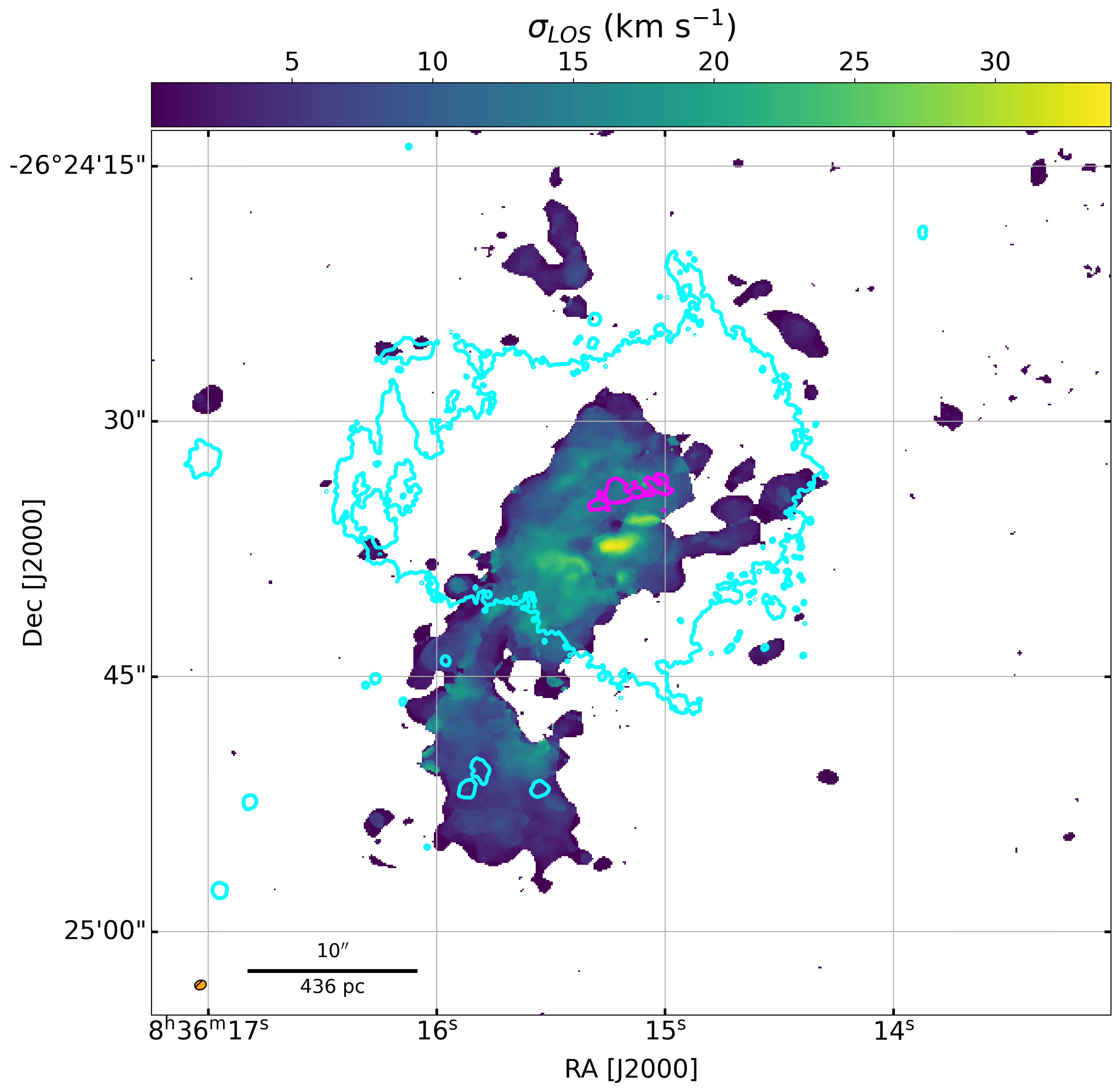}
    \label{fig:10mom2}}
    \caption{Combined \twco{(1-0)} maps of (a) the integrated intensity (b) intensity weighted velocity field, (c) the peak intensity, and (d) intensity weighted velocity dispersion, integrated over 780 -- 900~\kms. The raster maps are masked below 4$\sigma_{CO(1-0)}$. In each panel, the orange filled ellipse in the lower left is the combined \twco{(1-0)} beam. In (a) the nominal position of component D is outlined with red and blue circles sized to the OVRO and APEX beam, respectively. In (b) -- (d), the cyan contour is the H$\alpha$ ``footprint'', and in (b), isovelocity contours are shown in increments of 20~\kms. In (c) and (d), the fuchsia contour shows the VLA 33 GHz continuum from Figure \ref{fig:f555Finder}. In (d), the red dashed boxes outlines the regions shown in Figure \ref{fig:optical}.}
    \label{fig:CO10}
    
\end{figure*}

\begin{figure*}[htb!]
    \centering
    \subfloat{\includegraphics[width=0.42\linewidth]{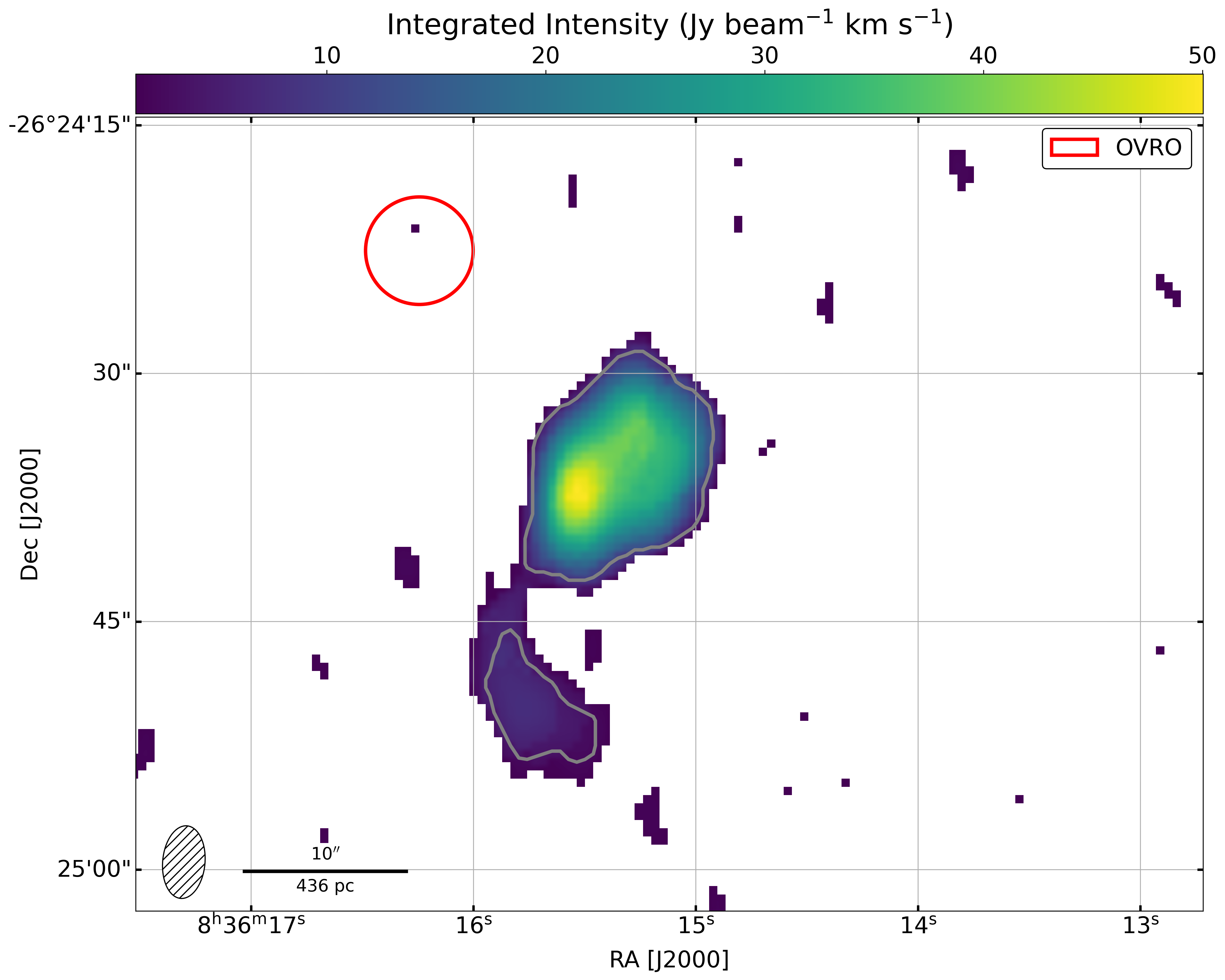}\label{fig:COmom0}}
    \subfloat{\includegraphics[width=0.42\linewidth]{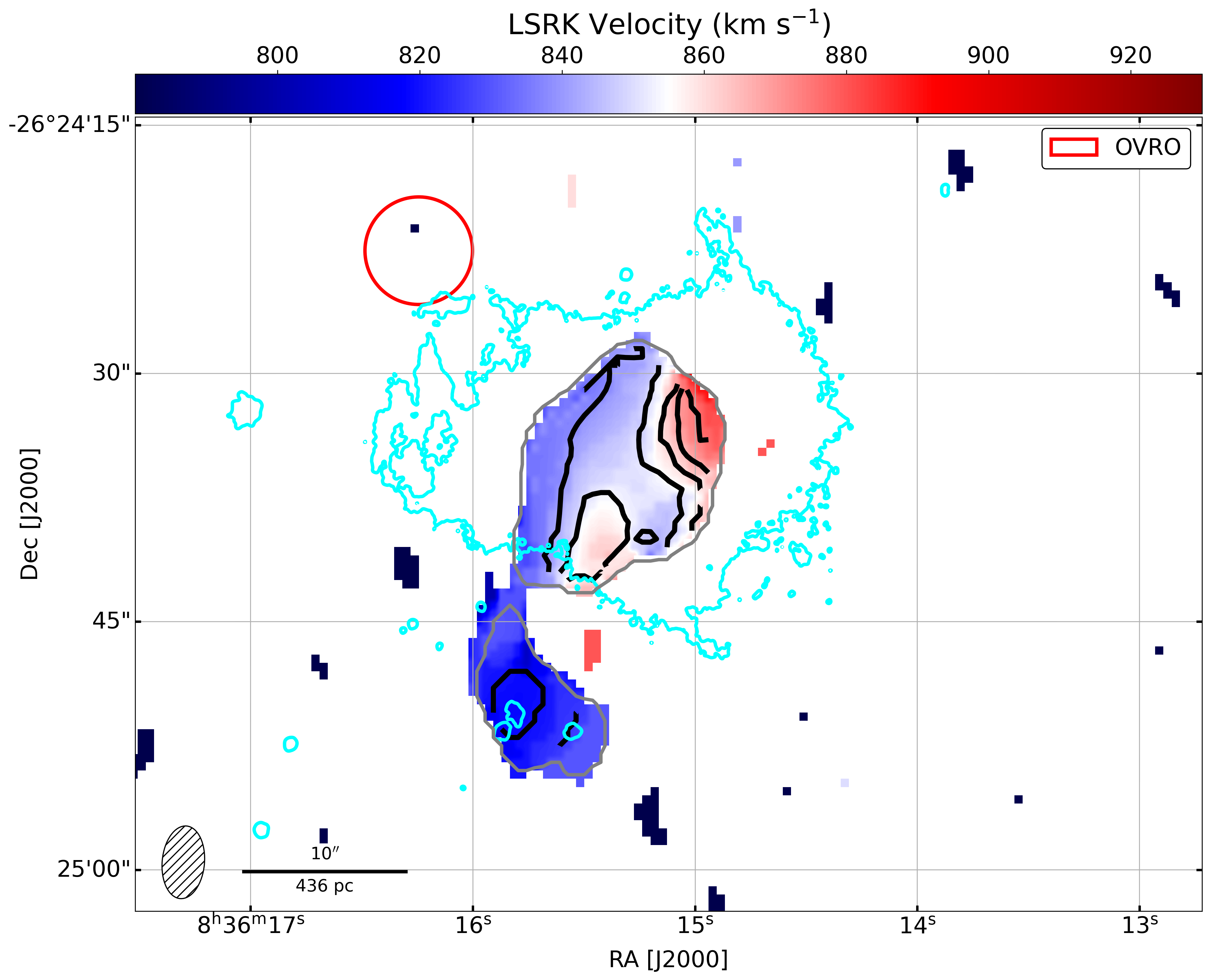}
    \label{fig:COmom1}}
    \quad
    \subfloat{\includegraphics[width=0.42\linewidth]{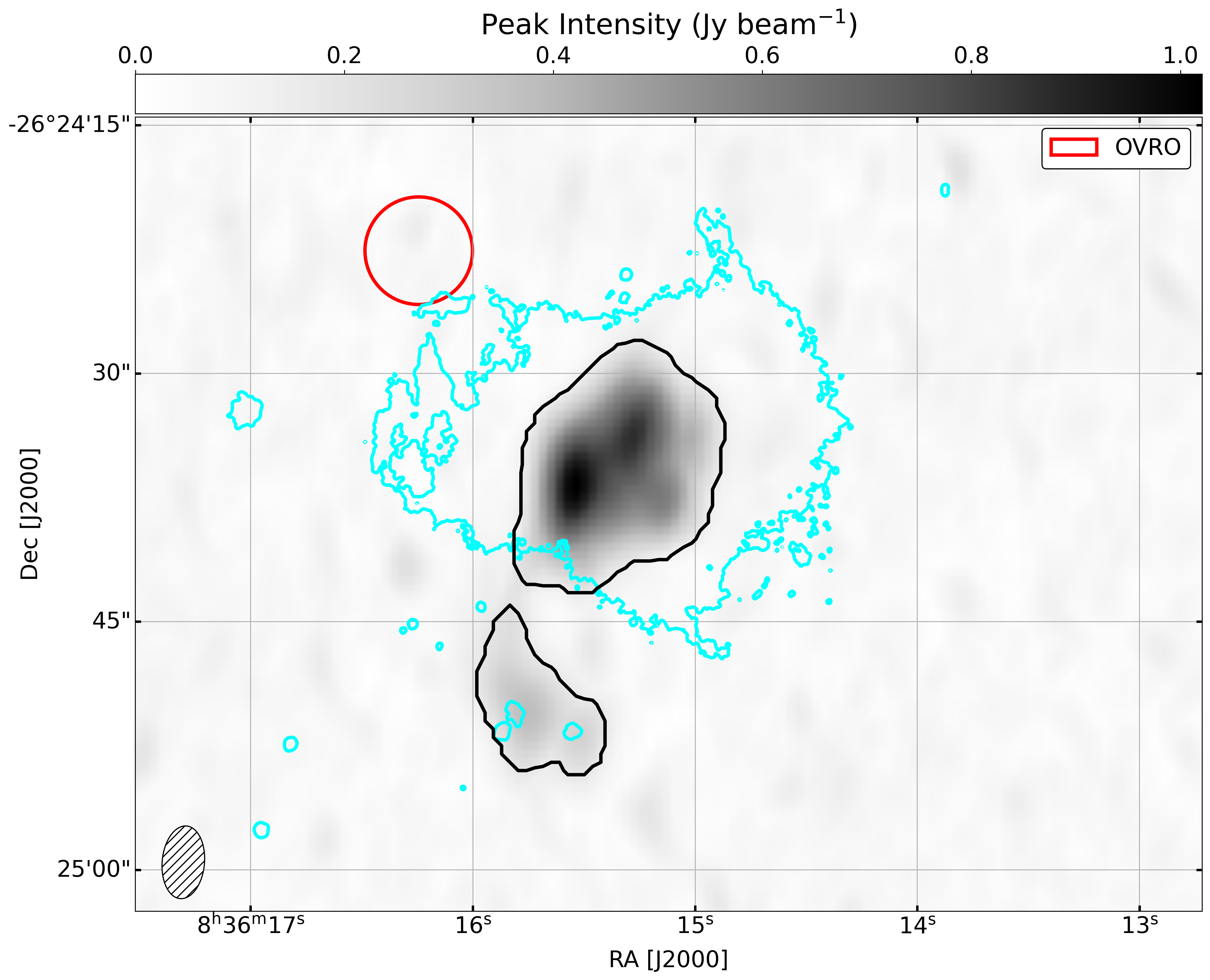}}
    \subfloat{\includegraphics[width=0.42\linewidth]{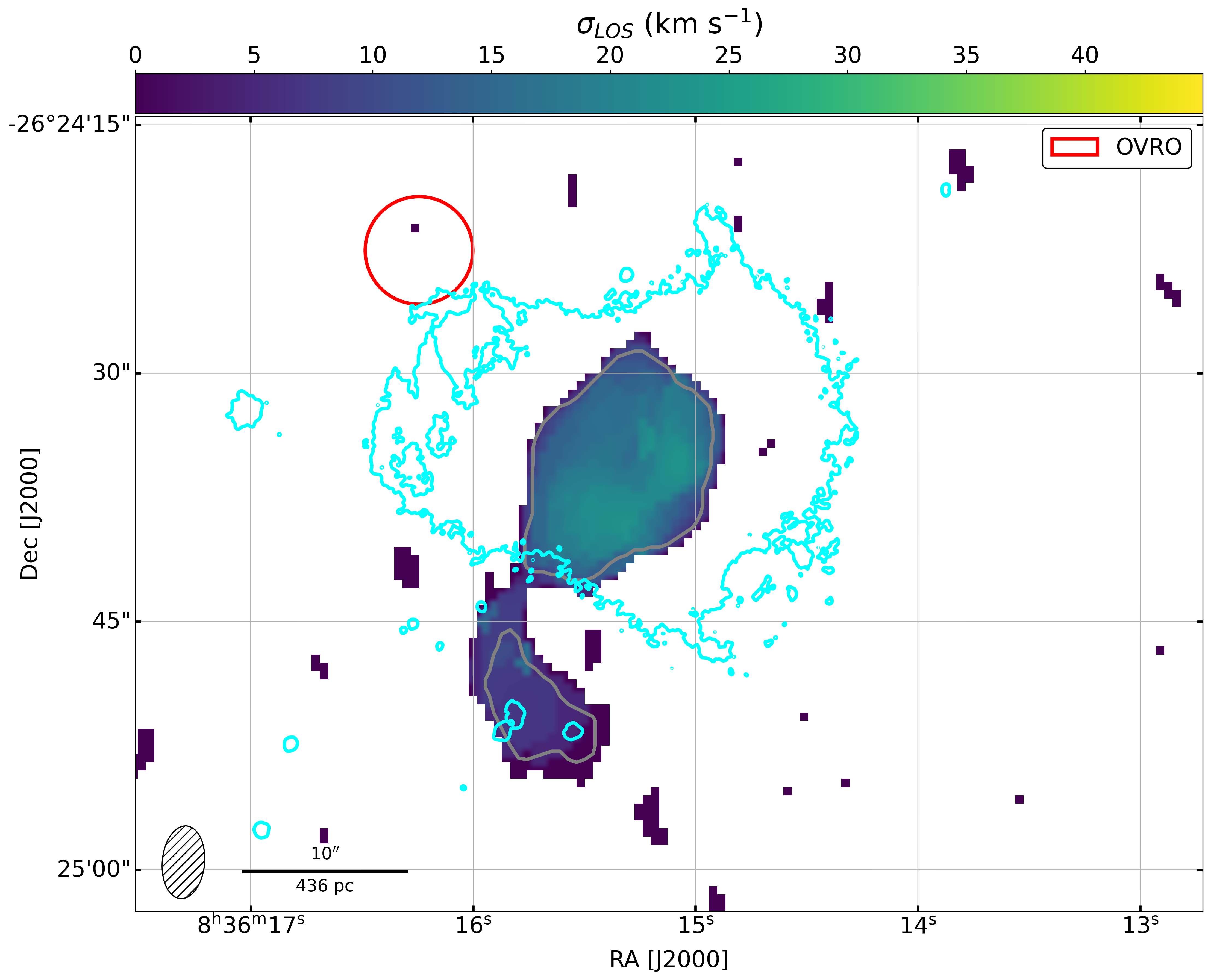}
    \label{fig:COmom8}}
    \caption{Combined \twco{(2-1)} maps of (a) the integrated intensity (b) intensity weighted velocity field, (c) the peak intensity, and (d) the intensity weighted velocity dispersion, integrated over 780 -- 940~\kms. The raster maps in (a) -- (c) are masked below 3$\sigma_{CO(2-1)}$. The nominal position of component D is outlined with red circle sized to the OVRO beam. The black contour shows 5$\sigma_{CO(2-1)}$ and the cyan contour is the H$\alpha$ ``footprint''. The black hatched ellipse in the lower left of each panel is the beam. In (b), isovelocity contours are shown in increments of 20~\kms.}
    \label{fig:COmom}
\end{figure*}

The CO emission in Hen 2-10 exhibits a pronounced southeast extension, clearly visible in both the \twco{(1-0)} and \twco{(2-1)} maps (Figures \ref{fig:CO10} and \ref{fig:COmom}, respectively). This extension corresponds to the CO ``tail'' first identified by \citetalias{Kobulnicky:1995}. The ``tail'' is notably offset from the galaxy’s central starburst regions and is blueshifted by approximately 20~\kms{} relative to the main CO body (Figure \ref{fig:coline}). The tail’s line width is also substantially narrower than that of the main body of the galaxy. While the main CO body is resolved into filamentary structures spanning $\sim$50~\kms{} in higher-resolution \twco{(3-2)} data \citep{Beck:2018}, the tail appears more kinematically coherent, with clumpy structures spanning only $\sim$20~\kms{} in the \twco{(1-0)} channel maps (Figure \ref{fig:CO10chan}; see also \citealt{Imara:2019}.

The morphology of the tail does not appear to result from spatial filtering; both the CARMA data presented here and the earlier OVRO observations are sensitive to arcminute-scale emission. Additionally, combining the CARMA \twco{(1-0)} data with the higher-resolution ALMA data recovers additional diffuse flux but does not eliminate the clumpy appearance of the tail, suggesting that the fragmentation is real and not an artifact of imaging. Although the \twco{(2-1)} observations have a larger beam and do not resolve individual clumps within the tail (see Figure \ref{fig:COchan}), the measured line widths and FWHM values are consistent with those derived from the higher-resolution \twco{(1-0)} data (Table \ref{tab:regions}). This agreement indicates that the bulk kinematics of the tail are reliably recovered in both transitions, even if the finer morphological details are beam-smeared in the lower-resolution maps.

The spatial relationship between star-forming regions and the molecular gas distribution reveals interesting relationships. In the \twco{(1-0)} peak intensity map (Figure \ref{fig:10mom8}), the two H$\alpha$ clumps known as Knots $\#$3 and $\#$4 \citep{Beck:1999,Mendez:1999} are visible at the southern end of the CO ``tail'', coinciding with regions of low CO emission, further supporting the interpretation that they are active star-forming sites physically associated with the CO ``tail''.

\begin{figure}[htb!]
    \centering
    \includegraphics[width=0.75\linewidth]{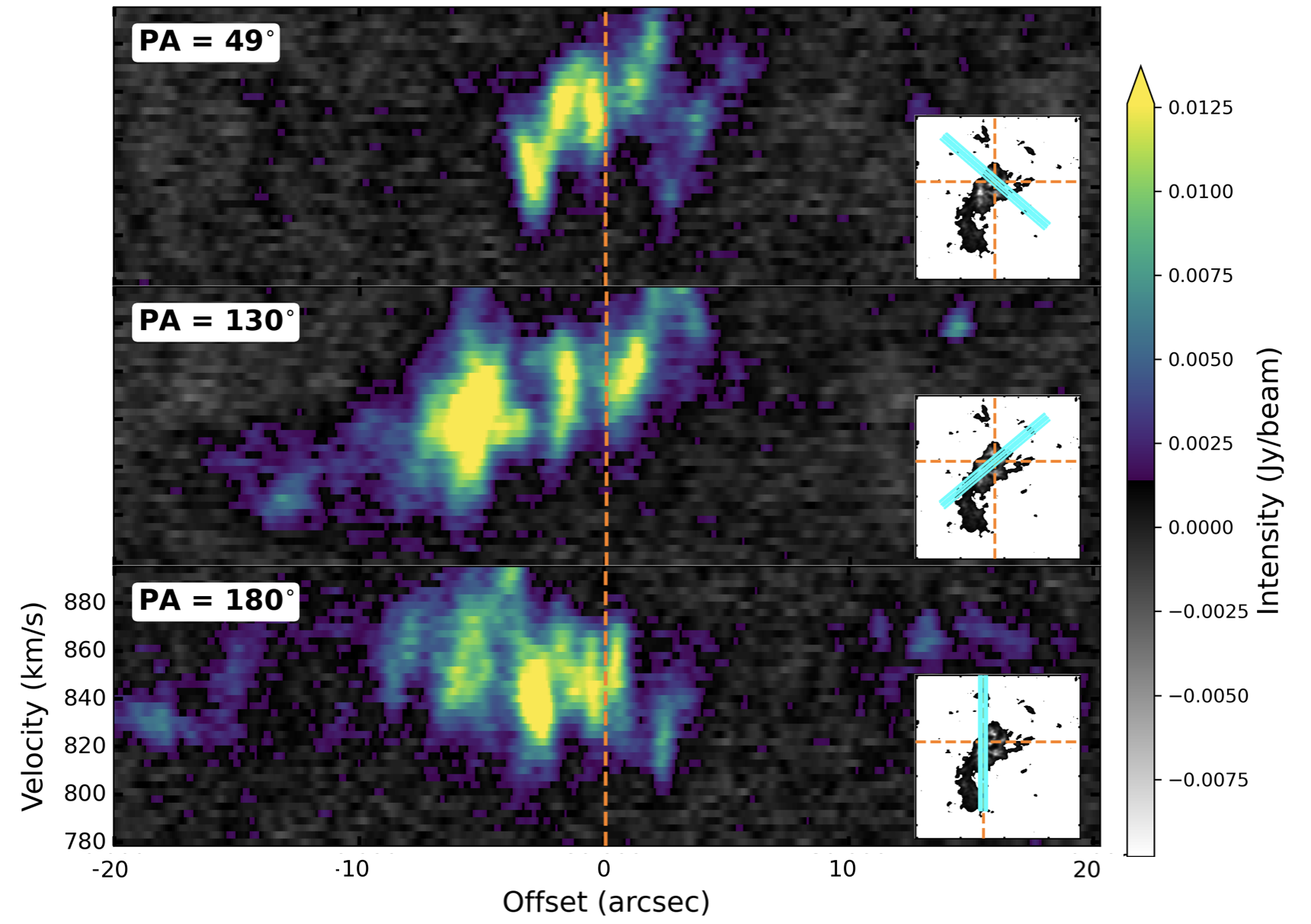}
    \caption{Same as Figure \ref{fig:pv_hi} but for the combined \twco{(1-0)} data. In the 180\ddeg{} slice, the slice is anchored at the midpoint between Regions A and B to intersect the `V-shaped' emission in the north as well as the southern tail.}
    \label{fig:COpv}
\end{figure}

There are two notable ($>$4$\sigma$ significance) clumps detected in our combined \twco{(1-0)} that have not been mentioned by previous authors but are worth including to construct a complete picture of the disturbed CO gas in \hen. Both have redshifted velocities near $\sim$870~\kms. The first is a distinct `V-shaped' structure directly north of Region A. The slice in the third panel of the \twco{(1-0)} position–velocity diagram (Figure \ref{fig:COpv}) intersects this emission, which shows this emission extending to the noise floor of the data. At that level, the feature appears kinematically detached from the main body. The second clump lies to the northwest of Region A and is intersected by the PA = 130\ddeg{} pv slice at (RA, Dec.) = (08:36:14.5, -26:24:23.8), which corresponds to the galaxy’s major kinematic axis as defined by \citetalias{Kobulnicky:1995}. This clump also appears unconnected to the bulk CO emission down to the 1$\sigma$ level. It has an integrated flux density of (0.63 $\pm$ 0.2) Jy~\kms, corresponding to a molecular gas mass of (0.8 $\pm$ 0.3) $\times$ 10$^6$ \Msun. Both clumps show narrow line profiles with FWHM $\sim$20~\kms. In the \twco{(2-1)} data, these structures are only tentatively detected at $<3\sigma$ significance and span less than a synthesized beam, making it likely that they are spurious features rather than real counterparts to the \twco{(1-0)} clumps. These clouds are on the edge of the primary beam of the \twco{(3-2)} data of \citet{Beck:2018}, but they are in the FOV of the \citet{Vanzi:2009} \twco{(3-2)} data. However, these features are not detected in those data, suggesting the clouds may be colder and thus under-luminous in the \twco{(3-2)} transition.

\citet{Vanzi:2009} report a detected $1.8 \times 10^7$ \Msun{} CO cloud to the NE of the starburst regions and suggest that it is potentially infalling to the galactic center. From their estimate, this cloud has a mass comparable to that of the CO ``tail'' (Section \ref{sec:comass})This estimated mass is comparable to that of the CO ``tail'' (Section \ref{sec:comass}), indicating the presence of two similarly massive CO structures situated in nearly orthogonal directions relative to the starburst regions. The fields of view of our combined \twco{(1--0)} and \twco{(2--1)} data encompass the region corresponding to component D (Figure \ref{fig:f555FinderB}). In the \twco{(1--0)} map, we find $<$4$\sigma_{CO}$ emission at 865~\kms{} within the beam of \citetalias{Kobulnicky:1995}, spatially coincident with the edge of the NE H$\alpha$ bubble. However, \citetalias{Kobulnicky:1995} and \citet{Vanzi:2009} find a velocity of +60~\kms{} from the systemic velocity, so this emission is not consistent with component D. No corresponding emission is seen in our \twco{(2--1)} data, which is more sensitive to warmer molecular gas. However, the \twco{(2--1)} observations lack the sensitivity required to detect this component, if it follows the brightness temperature ratios reported by \citet{Vanzi:2009}. Their \twco{(3--2)} detection peaks at 0.113 K; assuming a line ratio between the \twco{(3-2)} and \twco{(2-1)} transitions of $\sim$0.71, the expected \twco{(2--1)} brightness would be below our 1$\sigma_{CO(2-1)}$ sensitivity.

If the faint \twco{(1--0)} emission corresponds to component D, the integrated \twco{(1-0)} flux density of this feature is $\sim$99.6 mJy~\kms, corresponding to a molecular gas mass of $1.3 \times 10^5$ \Msun{} using Equation \ref{eq:mol}. This is nearly two orders of magnitude lower than the mass estimate based on the single-dish \twco{(3--2)} data. Spatial filtering is unlikely to explain the difference, since our interferometric data recover scales up to $\sim$115\arcsec. It is possible, however, that the 19\arcsec{} APEX beam used by \citet{Vanzi:2009} encompassed additional nearby CO emission. One candidate is the `V-shaped’ clump located within 15\arcsec{} of the reported position of component D. This feature has an integrated flux density of (1.8 $\pm$ 0.2) Jy~\kms, yielding a mass of $(2.3 \pm 0.7) \times 10^6$ \Msun, which is substantial but still well below the \citet{Vanzi:2009} value. As such, we cannot confidently identify this feature, nor identify other features, as component D.

\subsubsection{CO Mass\label{sec:comass}}

The total \twco{(1-0)} mass estimate is consistent with that reported by \citet{Imara:2019} and from single dish measurements of \citet{Baas:1994} and \citetalias{Kobulnicky:1995}. 
The total \twco{(2-1)} mass estimate is higher than previously reported mass of (3.9 $\pm$ 1.2) $\times$ 10$^7$ \Msun{} by \citet{Santangelo_Testi_Gregorini_Leurini_Vanzi_Walmsley_Wilner_2009}, which is expected given the combined CARMA and SMA data have better uv-coverage and sensitivity than either dataset alone. The mass estimates derived from the \twco{(1-0)} and \twco{(2-1)} transitions for both the main body and the total CO emission are consistent within the uncertainties, although the \twco{(2-1)}-based values are systematically lower. This difference likely reflects uncertainties in the adopted R$_{21}$ line ratio, which is not well constrained in Hen 2-10 and has known environmental dependencies. However, the mass estimates for the CO ``tail'' are not consistent between the two transitions using the same R$_{21}$ ratio as the main body. This R$_{21}$ ratio could underestimate the mass if the tail is colder. For example, adopting a lower line ratio of 0.6 would increase the inferred \twco{(2-1)} mass by approximately 38\%. Table \ref{tab:regions} lists the total molecular mass from the \twco{(1-0)} and the \twco{(2-1)} transitions for the three regions highlighted in the inset figure of Figure \ref{fig:coline}.

\section{Discussion \label{sec:discuss}}

Hen 2-10 is an intriguing galaxy: it is believed to host a low-luminosity black hole \citep{Reines:2011,Reines:2016} and it is undergoing an intense starburst episode with several nascent SSCs \citep{Johnson:2003} despite being a dwarf galaxy that appears isolated and without a companion. The cause of the starburst activity, and the numerous SSCs, has long been a source of speculation in the literature, with nature of the extended CO feature (``tail'') having a key role. In Hen 2-10, SSCs are forming near the interface between the kinematically distinct CO cloud, and the central region of the galaxy, suggesting the importance of the CO feature and the star formation. In a broader context, understanding the conditions favorable to SSC formation may provide insight into formation processes of globular clusters (GCs) in galaxies of the early universe \citep{Whitmore:1995,McLaughlin:2008,Longmore:2014,Kruijssen:2014,Lahen:2020,Horta:2021}.  Some have suggested that SSCs form preferentially in mergers, but it remains unclear whether this is a general conclusion, or one drawn largely from major and impressive mergers like the Antennae galaxies \citep[for a recent review, see][]{Bastian:2018}.

Here we consider several scenarios for the nature of the CO feature, and the extent to which existing observations support or refute these options: 
1) The CO feature is a result of material being expelled from the main body of the galaxy. 
2) The CO feature is due to previously expelled material falling back to the main body (i.e. a fountain). 
3) The CO feature is material falling into the main body of the galaxy for the first time. 
4) The CO feature is part of a nascent emerging spiral structure. 

\subsection{Could the CO ``tail'' be due to current or previous outflow?}
First we consider scenarios (1) and (2) and whether the extended CO feature might be due to expulsion of material either now or in the past by considering connections between the ionized gas and the molecular and neutral atomic gas. Internal mechanisms such as supernovae and stellar winds are potential drivers of episodic star formation, where feedback expels material that subsequently cools and re-accretes to trigger new star-forming events \citep{Stinson:2007}. In fact, the central star-forming regions in Hen 2-10 are known to be driving the NE and SW H$\alpha$ outflows (Figure \ref{fig:f555Finder}), and this material could break out of the galaxy or be recycled as a fountain \citep{Mendez:1999,Johnson:2000,Cresci:2017,Marasco:2023}. There is a 4$\sigma$ CO clump at the top of the NE H$\alpha$ bubble, which suggests a possible interaction between the outflow and molecular material (Figure \ref{fig:10mom1}). However, the CO material is only redshifted by +10~\kms{} relative to systemic velocity, whereas the H$\alpha$ bubble is +50 to +100~\kms{} from the systemic velocity. If the bubble is on the far side of the galaxy and expanding toward the far side of the galaxy, then it may be physically separated from this molecular emission despite their apparent alignment on the sky. This velocity discrepancy suggests that a direct interaction is unlikely, although the true three-dimensional geometry is uncertain.

If the CO ``tail'' were material expelled from the galaxy in a fountain and now in the process of re-accreting, we can estimate a lower bound for the cloud energetics by assuming that it is at apogee. The ``tail'' is $\sim$700 pc from the center of the galaxy. \citet{Reines:2011} estimate a stellar mass of $4 \times 10^9$ M$\odot$ within the inner few hundred parsecs, while nearly $10^{10}$ M$\odot$ is enclosed within 2 kpc \citep{Nguyen2014}. Assuming Keplerian motion, the orbital velocity at an apogee radius of 700 pc (equal to the semimajor axis in this approximation) is $\sim$135~\kms. Since the radial velocity of the cloud relative to systemic is small, this would imply the cloud lies within $\sim$20\ddeg{} of the plane of the sky (i.e., moving mostly in the plane of the sky).  Again assuming a semimajor axis of 700 pc about a central mass of 4 $\times$ 10$^9$ M$_{\odot}$, Kepler's third law dictates that the orbital half-period is 15 Myr. However, the kinetic energy required to launch the cloud out to 700 pc on this proposed orbit is 2 $\times$ 10$^{55}$ erg, or the energy of 2 $\times$ 10$^{4}$ supernovae divided by some efficiency factor rather less than 1. It seems unlikely nearly 10$^5$ supernovae exploded 15 Myr ago in the center of Hen 2-10, so an accreting/infalling scenario seems more plausible. 

The morphology and velocity dispersion of the CO ``tail'' are also not easily explained by an outflow scenario. The cold temperature (low R$_{32/10}$), narrow line widths ($\sim$20~\kms), physical confinement, and recent knots of star formation in the CO tail are difficult to reconcile with an outflow origin. Furthermore, the clear stellar outflows (i.e. the NE and SW bubbles) are nearly orthogonal to the CO ``tail''. Given the considerations discussed above, we consider the massive CO feature unlikely to be due to either current or previous outflow events. 

Evidence for a connection between the outflow and the neutral atomic gas may be present on larger scales. Although the resolution of the \hi{} data presented here limits our ability to probe near the nuclear H$\alpha$ emission (the H$\alpha$ emission is roughly contained within one \hi{} beam), there is a region with higher ($\sim$50~\kms) \hi{} velocity dispersion approximately 35\arcsec{} from the starburst regions that is roughly aligned, in the plane of the sky, with the NE H$\alpha$ bubble (Figure \ref{fig:HImom2}). Curiously, the highest-dispersion \hi{} gas lies just within the 1 \Msun{}~pc$^{-2}$ contour in Figure \ref{fig:HImom2}, which is a somewhat subjective but useful threshold for identifying galactic fountains \citep[e.g.,][]{McQuinn:2019}. This high-dispersion region may trace a previous outflow from the star-forming regions. If the material cannot escape the potential well of the galaxy, these higher \hi{} dispersions may mark the onset of a galactic fountain, in which gas expelled by feedback remains gravitationally bound and eventually re-accretes \citep[][]{Shapiro:1976,Chisholm:2016}, which could trigger future star formation. This alignment could be coincidental though, so higher resolution \hi{} data are needed to investigate this potential connection further.

\subsection{Could the CO ``tail'' be material infalling from the IGM?}

External mechanisms are also prevalent triggers for intense episodes of star formation.  \citetalias{Kobulnicky:1995} and \citet{Marquart:2007} suggest that Hen 2-10 is a moderately advanced merger, such that the progenitor galaxies are no longer distinct, but not so advanced that the molecular material has been completely accreted to the dynamic center. \citetalias{Kobulnicky:1995} suggest that the southern CO cloud is a tidal tail, resultant of such a merger/interaction. If Hen 2-10 has undergone a merger or interaction, then this CO material could be funneled to the center, which is triggering the star formation.

In the accreting/infalling case, one would expect a neutral hydrogen counterpart to the CO ``tail'', but as discussed previously, our current \hi{} data do not show any obvious features that can be linked spatially and kinematically to the CO ``tail''. Such \hi{} features can dissipate over time due to stellar feedback or over a few orbital periods \citep{Lelli:2014}. On the other hand, the low resolution of the \hi{} data may also be averaging over the features. Certainly, the \hi{} moment 1 and 2 maps exhibit disturbed kinematics (e.g., the velocity gradients in the S-SW; the large velocity dispersions). 

More sensitive and higher resolution \hi{} would offer insight into the merger or interaction history of Hen 2-10. For example, on the eastern outskirts of the galaxy, there may be an \hi{} feature that is intriguingly similar kinematically and spatially with the CO ``tail'': it appears as a beam-sized feature in the $\sim$770~\kms{} channel and joins with the bulk \hi{} gas by 810~\kms. The resolution of these new VLA data is too low to assess the morphology, and the higher resolution data of \citetalias{Kobulnicky:1995} do not extend to these velocity channels to investigate the structure further, though in their 813~\kms{} channel map, there is a suggestive \hi{} clump corresponding to this feature. As we cannot firmly rule out or confirm an \hi{} counterpart or an equally suggestive feature, we consider circumstantial evidence in favor of an infalling/accreting scenario.

\begin{figure}[htb!]
    \centering
    \subfloat{\includegraphics[width=0.52\textwidth]{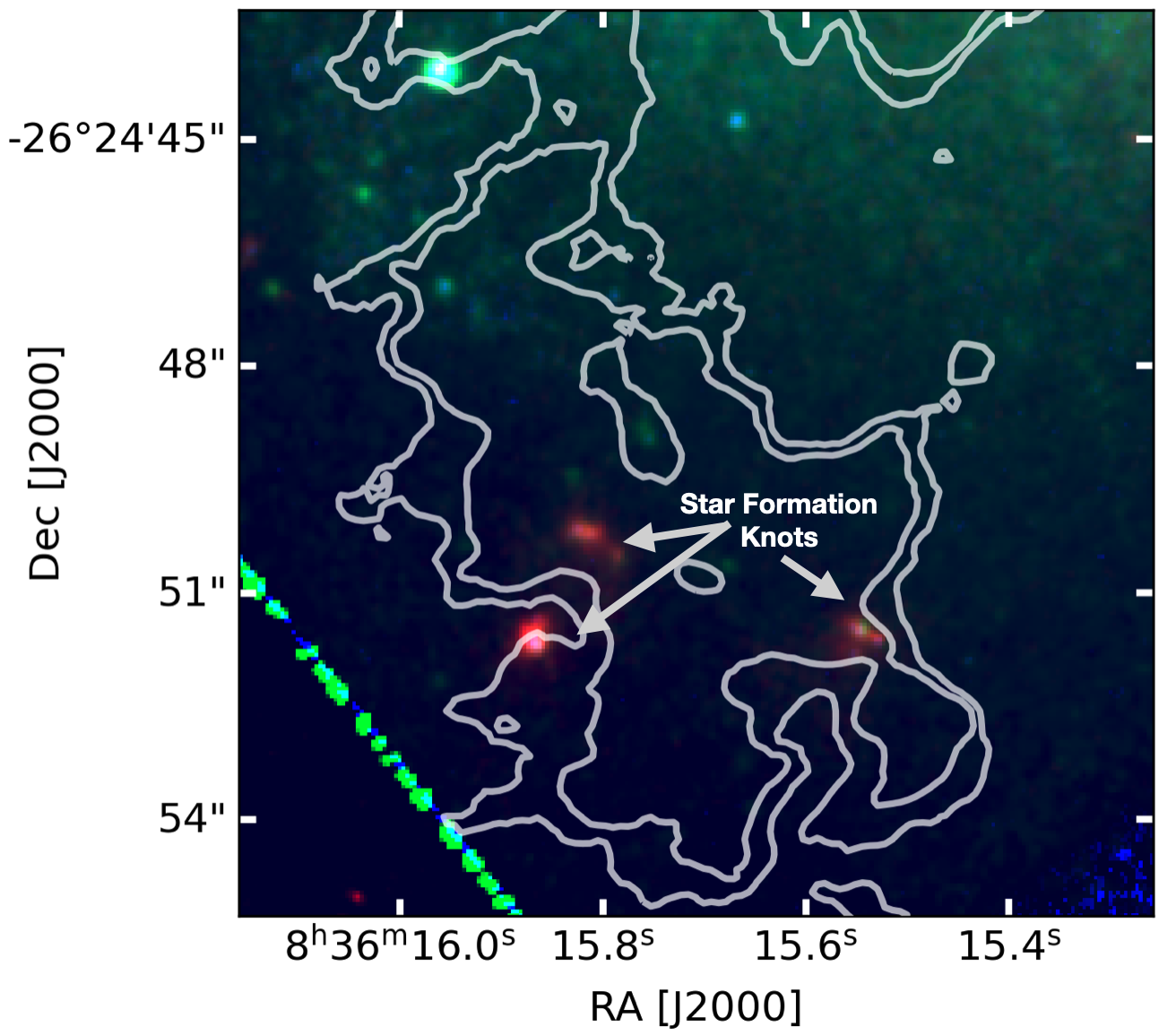}\label{fig:tricolorknots}}
    \subfloat{\includegraphics[width=0.43\textwidth]{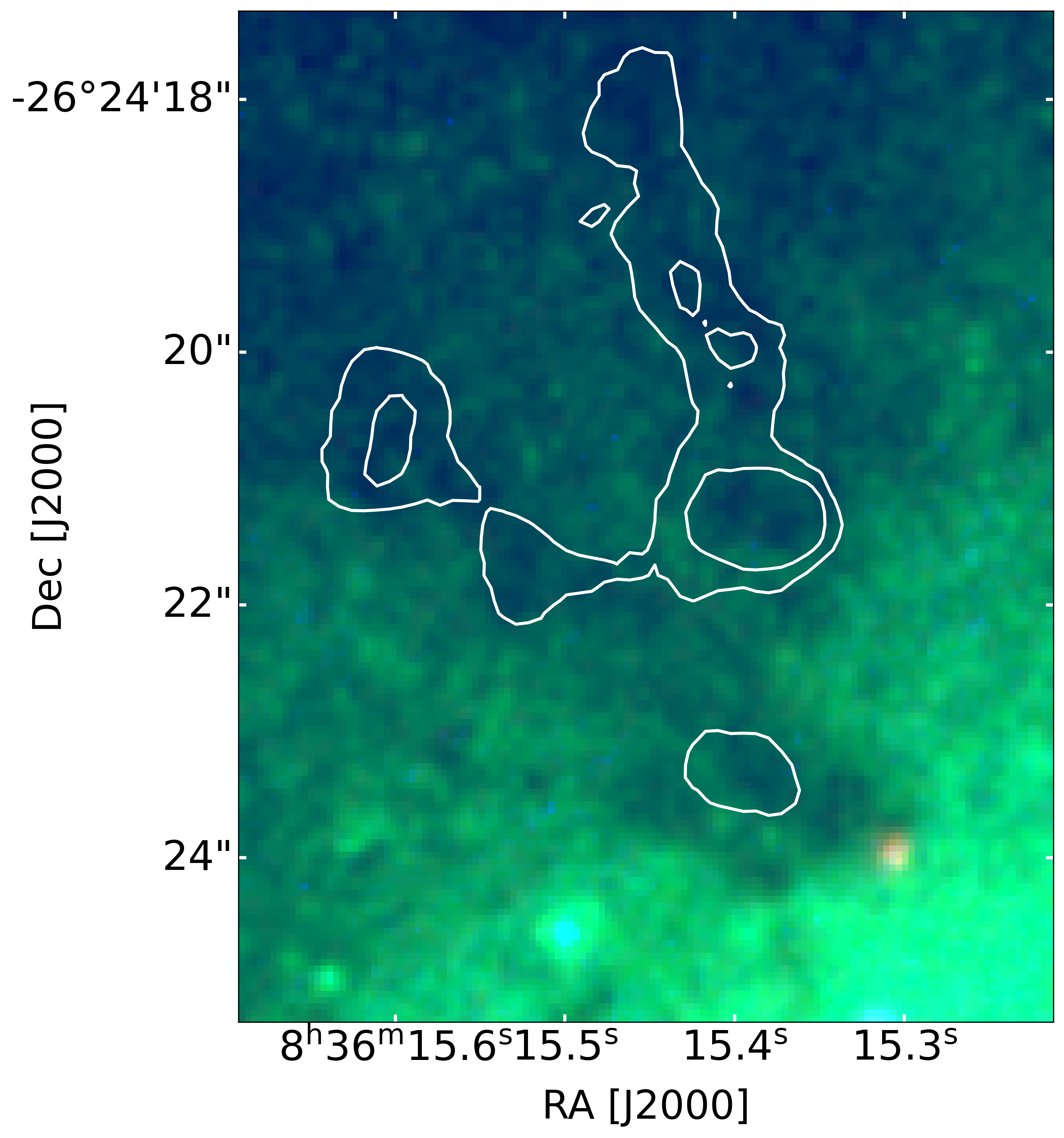}\label{fig:vshape}}
     \caption{The raster image is a three-color composite constructed from HST/WFPC2 data, where red corresponds to continuum-subtracted H$\alpha$, green to the F814W filter, and blue to the F555W filter. Note, the artifacts in green and blue are the edges of HST data.  The gray contours indicate \twco{(1–0)} moment 0 levels at 3 and 6 $\times;\sigma_{\mathrm{CO(1-0)}}$. The two panels provide zoomed-in views of regions outlined with dashed red boxes in Figure \ref{fig:10mom8}. Left: Close-up of the CO ``tail'', highlighting star-forming knots that are also prominent in H$\alpha$ emission. Right: Zoom-in on the `V-shaped' CO cloud, which corresponds to a dust extinction feature.
    }
    \label{fig:optical}
\end{figure}

The \twco{(1-0)} and the \twco{(2-1)} observations presented here suggest that the CO ``tail'' is colder than the CO associated with the main body of the galaxy. However, the CO ``tail'' is not entirely quiescent: it hosts star-forming clumps, including Knots $\#$3 and $\#$4 \citep{Beck:1999,Mendez:1999}, located at the southern end of the structure, far from the northern region where it connects to the main CO body (Figure \ref{fig:tricolorknots}). This offers additional circumstantial evidence that the CO feature is not likely to be due to outflow, but in fact may be the remnant of an extremely low-mass galaxy. On the other hand, the clouds at the northern end of the ``tail'', as outlined by the H$\alpha$ contour in Figure \ref{fig:f555FinderB}, appear to have sufficient mass to form SSCs, yet no such activity is observed \citep{Johnson:2018,Imara:2019}. It is not clear what the cause of the asymmetric star formation in this extended CO feature is.

Infrared observations also provide some guidance here by virtue of their ability to trace star formation. On galactic and kiloparsec scales, SFRs are known to correlate with \twco{(1-0)} luminosity \citep[e.g.,][]{bigiel08}. One of the most commonly used tracers of embedded star formation is 20 -- 25 $\mu$m emission, but the available Spitzer 24 $\mu$m data have insufficient angular resolution to resolve the CO ``tail'' from the main body. The 8 $\mu$m luminosity also correlates reasonably well with extinction-corrected H$\alpha$ and SFR (log L(8 $\mu$m) = log L(H$\alpha$)--1.67; \citealt{kennicutt09}), and Spitzer does resolve the main body and tail at 8 $\mu$m, with 210 $\pm$ 20 and 6.5 $\pm$ 1 MJy/sr, respectively (a ratio of 32). The ratio of CO intensity between the body and tail, from both the archival ALMA \twco{(1-0)} \citep{Imara:2019} and our CARMA data, is 3.2 $\pm$ 0.2. Thus, the SFR per CO intensity is \textit{10 times lower} in the CO ``tail'' than in the main body of the galaxy. This suppressed SFR further suggests that the molecular material in the tail is either at an earlier evolutionary stage, experiencing different environmental conditions, or possibly subject to a different CO-to-H$_2$ conversion factor, all of which support the interpretation that the tail is a recently accreted, still-settling structure rather than a product of past outflows.

It is important to note that there are other CO clouds potentially infalling/accreting onto the starburst regions. The CO clump and `V-shaped' cloud identified north/northwest of the main body do not appear to connect to the filamentary structures that have been proposed to fuel the central starburst. These northern clumps are compact, kinematically narrow, and spatially detached from the main CO body. With their lack of connection, the main CO body suggests they may represent a distinct population of molecular clouds, possibly in the early stages of accretion. The `V-shaped' CO cloud corresponds very well morphologically to dark features in the HST F555W image (Figure \ref{fig:vshape}), i.e. likely dust absorption. Thus, it is likely that it is in the front of the galaxy, and the large redshifts relative to the main body imply that it is infalling. 

Neither of the clumps are reported in the \twco{(3-2)} data of \citet{Vanzi:2009}, further suggesting that they may be cold. If so, they may evolve into more filamentary structures, like those seen in the sub-arcsecond \twco{(3-2)} maps of \citet{Beck:2018}, as they migrate inward and begin interacting with the turbulent, high-pressure environment of the central regions. These clumps could therefore provide a snapshot of an earlier phase in the inflow process, before the formation of coherent structures that channel gas to the starburst.

While the circumstantial evidence discussed above favors an accretion scenario, there are two considerations that are in tension with this conclusion. First, the velocity of the CO ``tail'' is blueshifted relative the main body of the galaxy. If the material is in front of the galaxy (as suggested by morphological correspondence with dark parts of the F555W image, i.e. potential dust lanes), an inflow scenario is geometrically implausible. If, however, the CO feature is infalling mostly transversely (in the plane of the sky and/or arcing around the main body of the galaxy), then some stars will be in front of the cloud and some will be behind.  The dark spots in the optical image that correspond to the CO ``tail'' are not as high contrast as those associated with the northern clouds, and it is entirely consistent with these observations that the gas is traveling mostly in the plane of the sky, obscuring only the fraction of the stars that are behind it, and displaying modest blueshifts relative to the center of the galaxy. Thus, we do not rule an infall scenario out based on this tension.

However, there is another critical problem to consider: if the CO feature is infalling from the intergalactic medium (IGM), where did it come from? While \hi{} clouds being accreting by galaxies are common, it is not clear how enriched molecular material on this scale would come about in the IGM in the absence of a previous generation of stars producing heavy elements. We consider this issue the most compelling argument against an accretion scenario, and it is not clear how it could be reconciled.

\subsection{Could Henize 2-10 have emerging spiral structure?}
\begin{figure}
    \centering
    \includegraphics[width=0.5\linewidth]{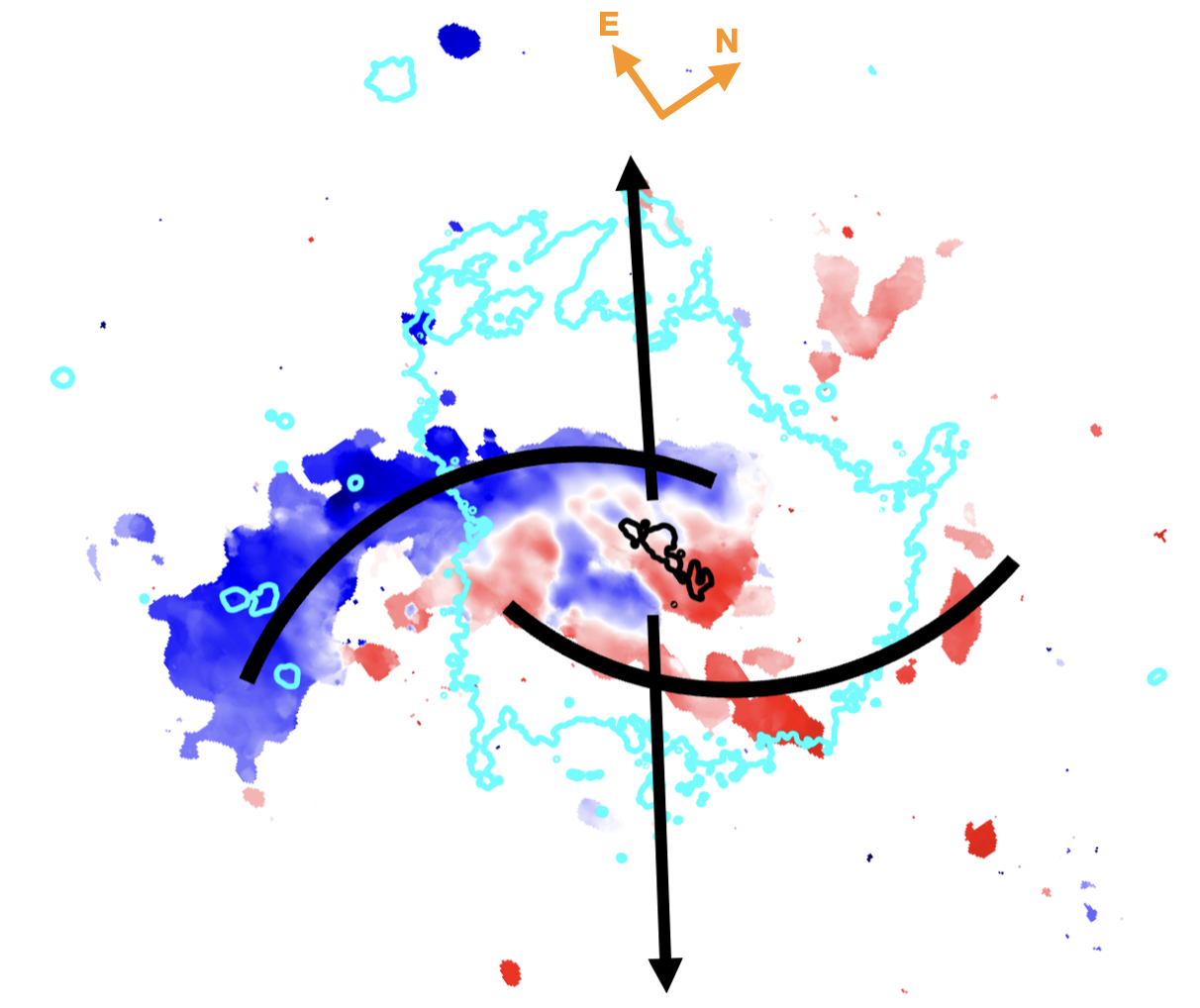}
    \caption{Cartoon diagram of spiral-like features overlain on a raster map of the \twco{(1-0)} intensity weight velocity field (see Figure \ref{fig:10mom1}). The raster map is rotated to emphasize the spiral. The cyan contour is the H$\alpha$ footprint, and the black contours are the VLA 33 GHz data, marking nuclear center of Hen 2-10 that hosts the AGN and SSCs. The black arrows are drawn to emphasize direction of the NE and SW outflows. The orange cardinal coordinates illustrate the Equatorial orientation. }
    \label{fig:spiral}
\end{figure}
Given the issues with each of the above scenarios, we also consider another option that might be at least as likely: perhaps Hen 2-10 is in the process of developing spiral structure. We add this scenario to the options for two main reasons: first, the velocity and emission structure of CO emission is suggestive of spiral structure (see Figure \ref{fig:spiral}). While the CO and \hi{} appear to be kinematically distinct, this has been observed in other spiral galaxies \citep[e.g.][]{Laudage:2024}. Hen 2-10 is also reminiscent of galaxies inferred to have emerging spiral structure in the CANDELS survey of galaxies at $1<z<3$, and would likely be classified as a ``clumpy spiral'' in the decision tree of  \citet{Margalef-Bentabol:2022}. We also note that Hen 2-10 is relatively massive for a dwarf galaxy; with \citet{Nguyen2014} estimating a $10^{10}$ \Msun{} enclosed within the inner 2 kpc,  the mass of Hen 2-10 overlaps with that of low mass spiral galaxies, and might be expected to be subject to more regularizing dynamical forces than lower mass galaxies. However, in the absence of more sensitive and higher resolution \hi{} observations, we only present this tentatively as an option. 

\subsection{Can an infalling/accreting CO cloud cause the starburst?}
If we assume the cloud is infalling or accreting, the next question is whether its impact could plausibly contribute to triggering intense star formation, like that seen in the center. To estimate an upper bound on its energy, we consider the most extreme scenario: infall from infinity. If the ``tail''  has fallen from infinity to 700 pc, it will reach 200~\kms{}. Since the line-of-sight velocity is only 25~\kms, such an infalling cloud would have to be traveling nearly within 10\ddeg{} of the plane of the sky.  From the observed CO surface brightness and the total mass of the ``tail'' calculated from CO (Table \ref{tab:regions}) distributed in a 150 pc radius sphere, we estimate a molecular number density of 50--80$\;$cm$^{-3}$. An impact of that gas traveling at 200~\kms{} will impart a ram pressure of P/k = 3--4 $\times$ 10$^8$ K$\;$cm$^{-3}$, well within the range suggested to trigger starburst activity and the formation of super star clusters \citep{Elmegreen:2002}. Such an infalling molecular cloud triggering the intense star formation activity seen in Hen 2-10 would add to our understanding of star-formation episodes in dwarf galaxies.

\section{Summary}
We investigate the distribution and kinematics of the molecular and neutral atomic gas to understand the cause of the intense star formation in the dwarf galaxy Henize 2-10. We present new VLA 21-cm observations as well as archival \twco{(1-0)} and \twco{(2-1)} observations. To investigate the molecular material, we combine \twco{(1-0)} observations taken with CARMA and ALMA; similarly, we combine \twco{(2-1)} observations from CARMA and SMA. The resulting maps offer improved sensitivity and image fidelity. The following is a summary of our results and conclusions:
\begin{itemize}
    \item The \hi{} maps (Figures \ref{fig:HImom0} \& \ref{fig:HImom8}) reveal an extended \hi{} envelope reaching over 2.5 kpc from the center of the galaxy, with a total \hi{} mass of (16.9 $\pm$ 0.5) \Msun. However, we do not find a striking or significant \hi{} tail, which was reported by \citetalias{Kobulnicky:1995}. Instead we find extended \hi{} features at many positions around the edge of the integrated intensity map, suggesting that there is diffuse extended emission below the threshold of these data. 
    \item The \hi{} intensity-weighted velocity and velocity dispersion maps (Figures \ref{fig:HImom1} \& \ref{fig:HImom2}) reveal disturbed kinematics. While the velocity field in the northern part of the galaxy shows a spider-like pattern characteristic of rotation, the gas in the south-southwest appears disordered. The spider-like morphology suggests the presence of a rotating disk, but the disk may be warped. The average \hi{} velocity dispersion, $\sigma_v$, is 8~\kms, consistent with typical values observed in the \hi{} distributions of dwarf galaxies \citep{Stil:2002}. However, two regions show significantly higher dispersions, with $\sigma_v > 50$~\kms; they are asymmetrically offset by $\sim$35\arcsec{} from the central starburst. 
    \item One of these \hi{} high-dispersion regions lies along the plane of the sky with the NE H$\alpha$ bubble, which is thought to be driven by stellar feedback \citep{Mendez:1999}. The region also falls within a contour corresponding to an \hi{} surface density of 1\Msun{}~pc$^{-2}$ (Figure \ref{fig:HImom2}). This suggests a possible connection between the \hi{} gas and the outflows from the starburst regions, raising the prospect of a developing galactic fountain or material escaping Hen 2-10 entirely.
    \item Though we cannot confirm a detection of the CO cloud known as component D, we identify two 5$\sigma$ CO clouds offset ($\sim$15\arcsec{}) from the bulk CO gas (Figure \ref{fig:CO10}). As these clouds have not been reported in higher J-transitions, it is possible that they are colder. We speculate they may represent a population of infalling/accreting clouds, though their origin is unknown.
    \item We also confirm the presence of a clumpy, kinematically coherent extension of molecular gas to the southeast, consistent with the previously reported CO ``tail'' from \citet{Kobulnicky:1995}. While the clumpy nature of the ``tail'' has been noted before in sub-arcsecond ALMA data, the inclusion of the CARMA data, which is sensitive to emission at larger scales, confirms the clumpy nature of the ``tail''. Although this material has an order of magnitude lower SFR per unit CO intensity than the gas near the starburst regions, it is forming stars.
    \item We explore whether the ``tail'' may be the result of an expulsion of material from the starburst, infalling/accreting due to a past merger or interaction, or associated with an emergent spiral structure. We favor the infalling/accreting scenario when considering the following circumstantial evidence:
    \begin{itemize}
        \item There are clear outflows (i.e. the NE and SW bubbles) that are nearly orthogonal to the CO ``tail'', so it would be unusual for an orthogonal outflow to exist as well.
        \item The ``tail'' has narrow linewidths and is coherent across its length, which would be difficult to reconcile in an outflow scenario.
        \item The southern end of the ``tail'' is forming stars while the material nearer to the interface with the main CO body have seemingly yet to form any. This asymmetry is difficult to understand for an outflow scenario but may be accommodated if the ``tail'' is a remnant of the merging companion galaxy.
        \item If this material is re-accreting due to a past outflow, such as in a fountain scenario, we estimate this amount of CO would require $\sim$10$^5$ supernovae to explode 15 Myr in the center of Hen 2-10, which seems unlikely.
    \end{itemize}
    \item If the CO ``tail'' is inflowing, we estimate it could impart enough pressure to plausibly contribute to starburst activity, such as that found at the center of Hen 2-10.

\end{itemize}

\newpage

\section{Acknowledgments}

We thank the anonymous referee for a cordial and collegial review. The thoughtful comments and suggestions greatly helped to refine the scope and clarity of this work.

This project was possible due to the support of the Post-Baccalaureate Student Research Program at the National Radio Astronomy Observatory. The National Radio Astronomy Observatory is a facility of the National Science Foundation operated under cooperative agreement by Associated Universities, Inc.

This paper makes use of the following ALMA data: ADS/JAO.ALMA\#2016.1.00027.S ALMA is a partnership of ESO (representing its member states), NSF (USA) and NINS (Japan), together with NRC (Canada), NSTC and ASIAA (Taiwan), and KASI (Republic of Korea), in cooperation with the Republic of Chile. The Joint ALMA Observatory is operated by ESO, AUI/NRAO and NAOJ.

The Hubble Legacy Archive (HLA) project augments the archival service provided for data from the Hubble Space Telescope (HST) and is a collaboration between the Space Telescope Science Institute (STScI/NASA), the Space Telescope European Coordinating Facility/European Space Agency (ST-ECF/ESA) and the Canadian Astronomy Data Centre/Canadian Space Agency (CADC/NRC/CSA). This research used the HLA facilities of the STScI, the ST-ECF and the CADC with the support of the following granting agencies: NASA/NSF, ESA, NRC, CSA.

\software{
    This research made use of the Common Astronomy Software Applications \citep[versions 4.2 \& 6.4;][]{CASA}, 
    the Cube Analysis and Rendering Tool for Astronomy \citep{CARTA},
    Astropy \citep[version 5.3.4;][]{Astropy:2013,Astropy:2018},
    Photutils \citep[version 1.11.0;][]{Bradley:1.11},
    Spectral-Cube \citep[version 0.6.5;][]{Ginsburg2019},
    Pvextractor (version 0.3; \url{https://pvextractor.readthedocs.io/en/latest/}),
    NumPy \citep[version 1.26.4 ][]{numpy2020},
    SciPy \citep[version 1.13.1][]{scipy}, and
    Matplotlib \citep[version 3.9.1][]{matplotlib}.
}

\appendix
\section{Channel Maps\label{app:chan}}
\begin{figure*}[htb!]
    \centering
    \includegraphics[width=1\linewidth]{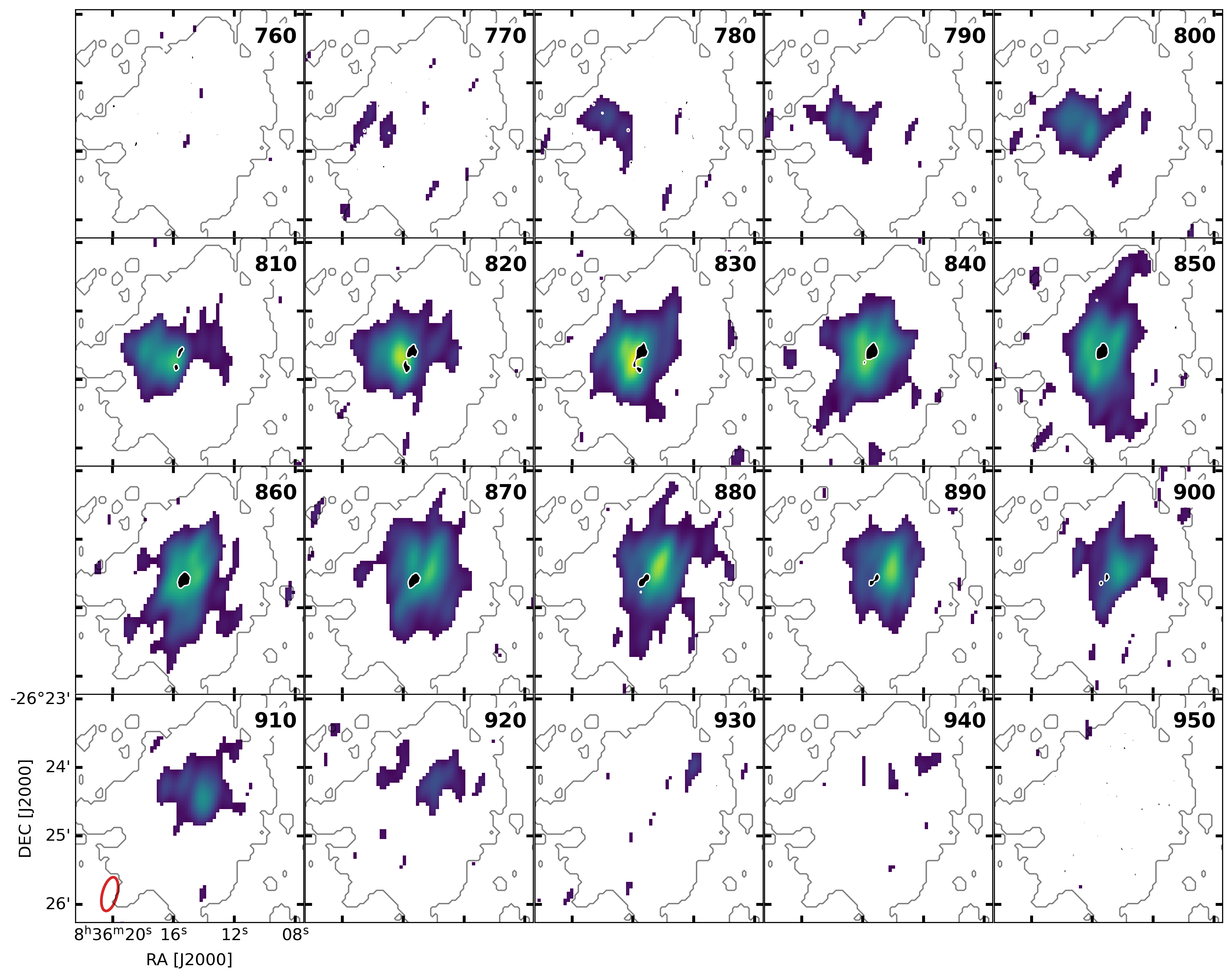}
    \caption{\hi{} channel maps in increments of 10 \kms; each figure is labeled with the velocity (\kms) in the LSRK frame. The \hi{} data in the red-yellow-blue raster are masked, per channel, below 3$\sigma_{HI}$ and share a common scale. The black contour shows the \hi{} moment 0, masked below 3$\sigma_{HI}$ and integrated over 730 -- 990 \kms. The \hi{} beam is shown with a black hatched ellipse in the lower left of the 900 \kms{} panel. Overlaid on the \hi{} raster are the channel matched combined \twco{(2-1)} 5$\sigma_{CO}$ data, shown as a filled black contour (see Figure \ref{fig:COchan}).}
    \label{fig:HIchan}
\end{figure*}

\begin{figure*}[htb!]
    \centering
    \includegraphics[width=1\linewidth]{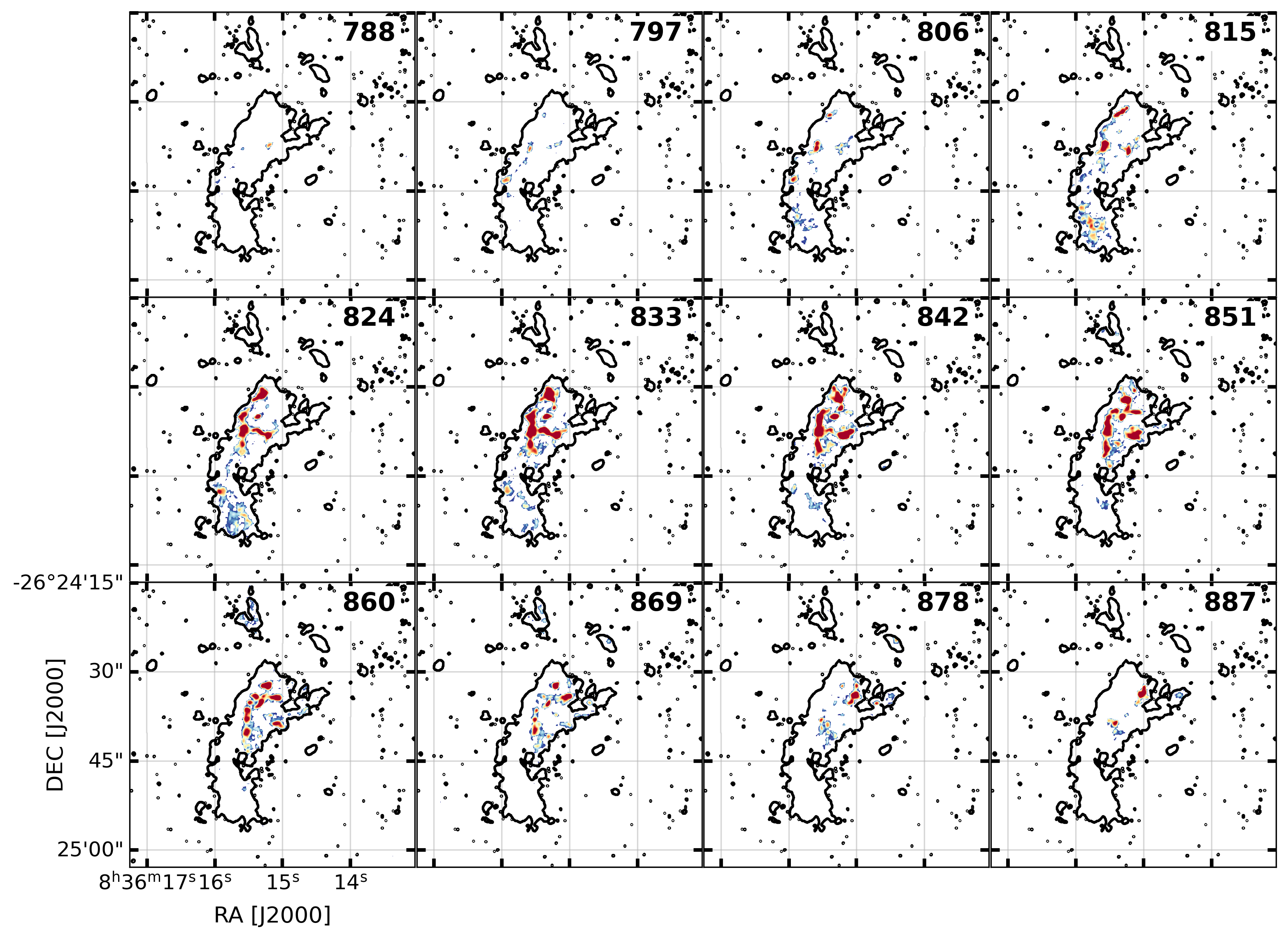}
    \caption{Combined CARMA + ALMA \twco{(1-0)} channel maps in increments of 9 \kms; each figure is labeled with the velocity (\kms) in the LSRK frame. The raster data are masked per channel below 4$\sigma_{CO}$ and share a common raster scale. The black contour shows the integrated \twco{(1-0)}, masked below 4$\sigma_{CO}$. The \twco{(1-0)} beam is shown with a black ellipse in the lower left of the 860~\kms{} panel.}
    \label{fig:CO10chan}
\end{figure*}

\begin{figure*}[htb!]
    \centering
    \includegraphics[width=0.9\linewidth]{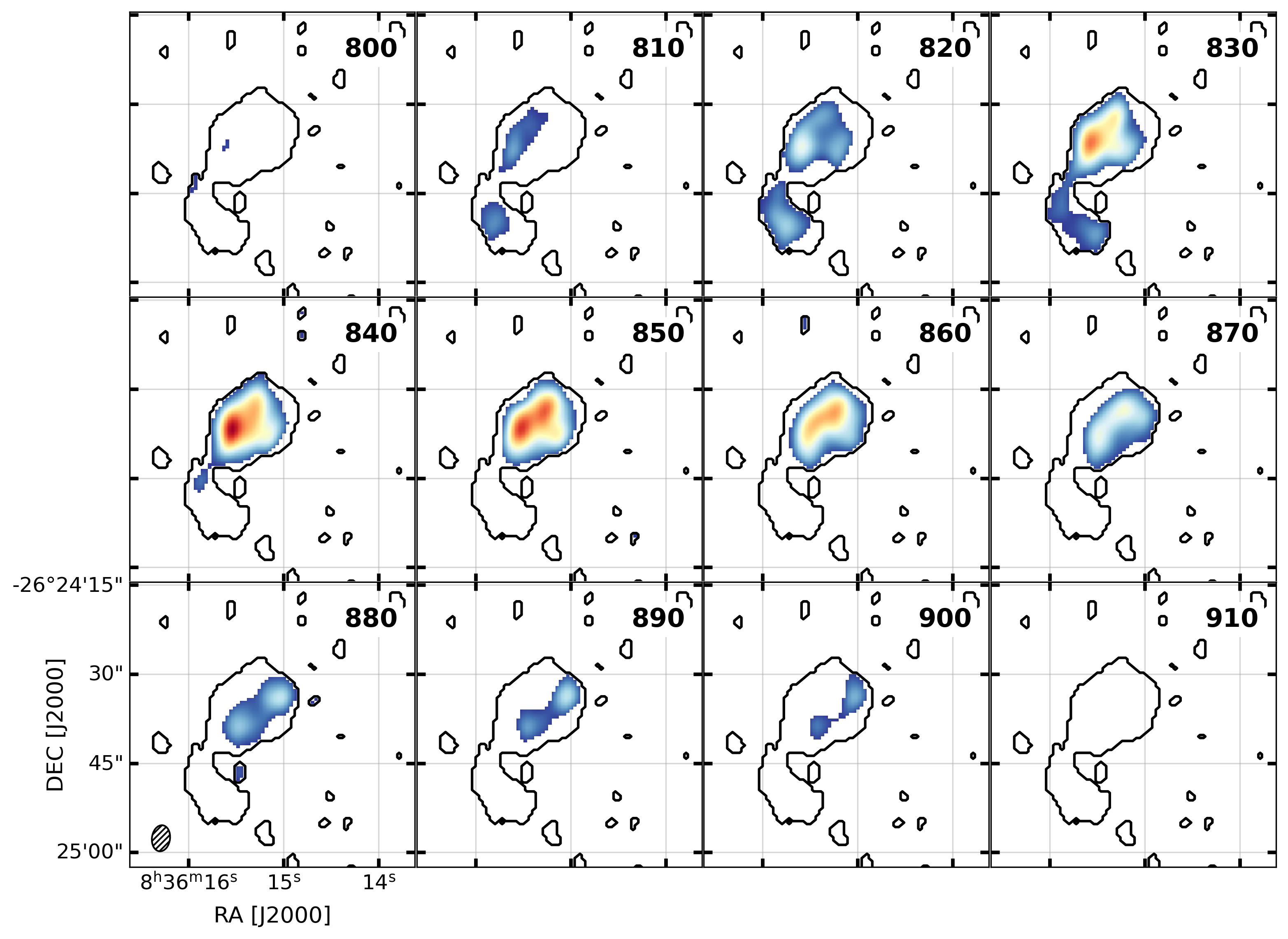}
    \caption{Combined CARMA + SMA \twco{(2-1)} channel maps in increments of 10 \kms; each figure is labeled with the velocity (\kms) in the LSRK frame. The raster data are masked per channel below 5$\sigma_{CO}$ and share a common raster scale. The black contour shows the integrated \twco{(2-1)}, masked below 5$\sigma_{CO}$ and integrated over 730 -- 990~\kms{} to match the \hi{} channel map. This velocity range is wider than that from Figure \ref{fig:COmom0}. The \twco{(2-1)} beam is shown with a black hatched ellipse in the lower left of the 900~\kms{} panel.}
    \label{fig:COchan}
\end{figure*}

The channel maps for the \hi{}, \twco{(1-0)}, and \twco{(2-1)} data sets are presented here in Figures \ref{fig:HIchan}, \ref{fig:CO10chan}, and \ref{fig:COchan}, respectively.

\newpage
\clearpage

\newpage
\clearpage

\bibliography{Henize2-10,Antennaebib}{}
\bibliographystyle{aasjournalv7.bst}

\end{document}